\documentclass[12pt,preprint]{aastex}

\lefthead{K. Krisciunas {\em et al.}}
\righthead{Supernova 2001el}

\begin{document}

\title{Optical and Infrared Photometry
 of the Nearby Type Ia Supernova 2001el}
\author{Kevin Krisciunas,$^1$
Nicholas B. Suntzeff,$^1$
Pablo Candia,$^1$
Jos\'{e} Arenas,$^1$
Juan Espinoza,$^1$
David Gonzalez,$^1$
Sergio Gonzalez,$^2$
Peter A. H\"{o}flich$^3$,
Arlo U. Landolt,$^4$\altaffilmark{,6}
Mark M. Phillips,$^2$ and 
Sergio Pizarro$^1$}
\affil{$^1$Cerro Tololo Inter-American Observatory, National Optical
Astronomy \\ Observatories,\altaffilmark{5} Casilla 603, La Serena, Chile \\
$^2$Las Campanas Observatory, Carnegie Observatories, Casilla 601, La
Serena, Chile \\
$^3$University of Texas, Dept. of Astronomy, Austin, TX 78712 \\
$^4$Louisiana State University, Dept. of Physics and Astronomy, Baton
Rouge, LA 70803-4001 }
\altaffiltext{5}{The National Optical Astronomy
Observatories are operated by the Association of Universities for
Research in Astronomy, Inc., under cooperative agreement with the
National Science Foundation.} 
\altaffiltext{6}{Visiting Astronomer, Cerro Tololo
Inter-American Observatory, National Optical Astronomy Observatories.}

\email {kkrisciunas, nsuntzeff@noao.edu \\
pcandia, juan, spizarro, dgonzalez@ctio.noao.edu \\
sergiogonzalezctio@yahoo.com \\
jose@gemini.cfm.udec.cl \\
pah@astro.as.utexas.edu \\
mmp@lco.cl \\
landolt@baton.phys.lsu.edu }

\begin{abstract} 

We present well sampled optical ($UBVRI$) and infrared ($JHK$) light curves
of the nearby ($\approx$18.0 Mpc) Type Ia supernova SN 2001el, from 11 days
before to 142 days after the time of $B$-band maximum.  The data represent
one of the best sets of optical and infrared photometry ever obtained for a
Type Ia supernova.  Based on synthetic photometry using optical spectra of SN
2001el and optical and infrared spectra of SN 1999ee, we were able to devise
filter corrections for the $BVJHK$ photometry of SN 2001el which to some
extent resolve systematic differences between SN 2001el datasets obtained
with different telescope/filter/instrument combinations.  We also calculated
$V$ minus infrared color curves on the basis of a delayed detonation model
and showed that the theoretical color curves match the unreddened loci for
Type Ia SNe with mid-range decline rates to within 0\fm2. Given the
completeness of the light curves and the elimination of filter-oriented
systematic errors to some degree, the data presented here will be useful for
the construction of photometric templates, especially in the infrared. On the
whole the photometric behavior of SN 2001el was quite normal.  The second
$H$-band maximum being brighter than the first $H$-band maximum is in accord
with the prediction of Krisciunas et al. (2000) for Type Ia SNe with
mid-range decline rates.  The photometry exhibits non-zero host extinction,
with total A$_V$ = 0.57 $\pm$ 0.05 mag along the line of sight.  NGC 1448,
the host of SN 2001el, would be an excellent target for a distance
determination using Cepheids.

\end{abstract}

\keywords{supernovae, photometry; supernovae}

\section{Introduction}

Type Ia supernovae (SNe) continue to be some of the most important objects
of study in extragalactic astronomy.  Since the discovery of the
luminosity-decline rate relation (Phillips 1993; Phillips et al. 1999) and
the corresponding relationships between light curve shapes and
luminosities-at-maximum (Riess, Press, \& Kirshner 1996; Riess et al. 1998;
Perlmutter et al. 1997), Type Ia SNe have been regarded as standardizable
candles to be used for calibrating the distances to the host galaxies.

Until recently, the amount of published infrared data
of Type Ia SNe was quite small (Elias et al. 1981, 1985; 
Meikle 2000, and references therein; Krisciunas et al. 2000, 2001). 
Beginning in 1999 two of us (MMP and NBS) began a coordinated effort,
using telescopes at Las Campanas and Cerro Tololo, to obtain well-sampled
optical and infrared light curves of Type Ia supernovae.  The first of
these, SN 2000bk, was published by Krisciunas et al. (2001), including
data from Apache Point Observatory.  

Recently, we have published well-sampled light curves of SNe 1999aw 
(Strolger et al. 2002) and 1999ac (Phillips et al. 2002).  In  
the $H$- and $K$-bands, in particular, we note that the light curves 
of Type Ia SNe are quite flat 
about 10 days after the time of $B$-band maximum, T($B_{max}$), allowing
us get an accurate brightness measure of Type Ia SNe in the near-infrared,
independent of the peculiarities of the light curves around the time of
the primary maximum ($t \approx -3$ d) or the time
of the secondary maximum ($t \approx +25$ d).\footnote[7]{As
is conventional, we shall measure the ``age'' of a supernova in time
before or after T($B_{max}$).  For this we use the variable $t$, measured
in days.}

In addition to the data of SN 2001el presented here, we are finishing up the
reduction of the light curves of another dozen objects. This will allow us to
characterize better than ever before the infrared light curves of Type Ia SNe and
correlate the variations observed at infrared wavelengths with those seen at
optical wavelengths.  Because interstellar extinction at infrared wavelengths is
an order of magnitude less than extinction at optical wavelengths, we expect that
we will be able to exploit this fact to obtain intrinsic photometry of Type Ia
SNe only minimally affected by interstellar extinction.  This will lead to an
infrared Hubble diagram for Type Ia SNe free of significant systematic errors
incurred from imprecise reddening estimates, and characterized by small random
errors. Only with such a foundation for {\em nearby} Type Ia SNe can we have
confidence in the conclusions drawn from observations of {\em distant} SNe, such
as evidence for a non-zero cosmological constant (Riess et al. 1998, Perlmutter
et al. 1999).

\section{Observations}

SN 2001el was discovered by Monard (2001) on 17.064 September 2001 UT (=
Julian Date 2,452,169.564) some 22\arcsec\ west and 19\arcsec\ north of the
nucleus of the SAcd galaxy NGC 1448, which incidentally was also the host of
the Type II SN 1983S. According to the NASA/IPAC Extragalactic Database (NED)
the heliocentric velocity of this galaxy is 1164 km s$^{-1}$. Based on a
spectrum taken with the VLT on 21 September 2001 UT (JD = 2,452,173.5),
Sollerman, Leibundgut, \& Lundqvist (2001) showed SN 2001el to be a Type Ia SN well
before maximum.  On this basis we requested that frequent optical and
infrared imagery be made with the Yalo-AURA-Lisbon-Ohio (YALO) 1-m telescope
at Cerro Tololo.  We also obtained optical imagery on five nights with the
CTIO 0.9-m telescope.  Optical data of Landolt presented here is
single-channel photoelectric photometry carried out with the CTIO 1.5-m
telescope.  Finally, we obtained other infrared imagery with the 1-m Swope
telescope at Las Campanas, which was used, along with observations of
infrared standards of Persson et al. (1998) to calibrate the $JHK$ magnitudes
of the field stars near SN 2001el.

Fig. 1 is a $V$-band optical image in which we identify the location of SN
2001el and various local field standards.  In Tables 1 and 2 we give optical and
infrared photometry of some of these local standards.  These were used as
secondary standards tied to the optical and infrared standards of Landolt
(1992) and Persson et al. (1998), respectively.  This allowed us to calibrate
the photometry of the SN on photometric as well as non-photometric nights.

Photometry reduction of the YALO data was carried out using {\sc daophot}
(Stetson 1987, 1990) and a set of programs written by one of us (NBS).  
The photometric techniques were described by Suntzeff et al. (1999) and
involve using aperture magnitudes and aperture corrections for the brighter
stars, while relying on point spread function (PSF) magnitudes for fainter stars.  
We used transformation equations of the following form:

\begin{equation}
b \; = B \; + \; ZP_b \; - \; 0.079 (B-V) \; + k_b X \; ;
\end{equation}

\begin{equation}
v \; = V \; + \; ZP_v \; + \; 0.018 (B-V) \; + k_v X \; ;
\end{equation}

\begin{equation}
r \; = R \; + \; ZP_r \; - \; 0.030 (V-R) \; + k_r X \; ;
\end{equation}

\begin{equation}
i \; = I \; + \; ZP_i \; + \; 0.045 (V-I) \; + k_i X \; ;
\end{equation}

\begin{equation}
j \; = J \; + \; ZP_j \; - \; 0.034 (J-H) \; + k_j X \; ;
\end{equation}

\begin{equation}
h \; = H \; + \; ZP_h \; + \; 0.022 (J-H) \; + k_h X \; ;
\end{equation}

\begin{equation}
k \; = K \; + \; ZP_k \; - \; 0.003 (J-K) \; + k_k X \; .
\end{equation}

\parindent= 0 mm

The parameters on the left hand sides are the instrumental magnitudes. On the right
hand side of each equation we have the standardized magnitude, a photometric
zeropoint, a color term times a standardized color, and an extinction term times the
air mass.  Given that the color terms are reasonably close to zero, the YALO filters
provide a good, but not perfect, match to the filters used by Landolt (1992) and
Persson et al. (1998) for the derivation of their standard star networks.

\parindent= 9 mm

In Table 3 we present $UBVRI$ photometry of SN 2001el.  A majority of the
data is from YALO, calibrated using the secondary standards from Table 1.
Landolt's data are tied directly to the Landolt (1992) standards, i.e.
without using the secondary standards.  The internal errors of the
Landolt data are as small as several millimags.  His data were taken when
the SN was getting monotonically fainter, and the data bear this out on
those four nights when he measured the SN on more than one occasion.  
For our field star number 1 Landolt obtained mean values of $V, B-V,
V-R$, and $V-I$ which are +0.003, +0.002, $-$0.005, and $-$0.009 mag
different, respectively, than the values obtained from CCD photometry
with YALO.  Thus, photometry of {\em stars} carried out with the CTIO
1.5-m and YALO telescopes is in excellent agreement.

We note that Landolt's measurements of SN 2001el are aperture photometry
using a 14 arcsec diameter aperture.  His sky readings were taken 15 arcsec
to the north of supernova.  To determine the amount of diffuse light of
NGC 1448 which was in his diaphragm while measuring SN 2001el, we used
CCD frames of the field taken 16 October 2001 and 15 January 2002.  We
determined that $U$-band corrections to Landolt's photoelectric photometry
are negligible, 0\fm002 or less.  In $B$, $V$, $R$, and $I$, however,
the data required correction by 0\fm019 to 0\fm043, 0\fm024 to 0\fm035,
0\fm034 to 0\fm048, and 0\fm058 to 0\fm078, respectively.  (Another
way of describing this is that galaxy light is {\em red}.) The Landolt
data in Table 3 include these corrections.

There were four nights when Landolt observed SN 2001el when it was
also observed with YALO. In the sense ``YALO {\em minus} Landolt''
the $B$-band data show a difference of +0\fm037.  For other bands
$\Delta V$ = $-$0\fm025, $\Delta R$ = $-$0\fm009, and
$\Delta I$ = $-$0\fm064.  $V$-band data obtained with the CTIO 0.9-m
telescope by Arenas is up to 0.067 mag fainter at the time of maximum light
than corresponding YALO data.

Suntzeff (2000) notes that one observes sizable and repeatable systematic
differences for the $B-V$ colors of Type Ia SNe measured with different
telescopes.  In the cases of SNe 1998bu and 1999ee observed with the YALO
and CTIO 0.9-m telescopes, one finds that YALO colors are 0\fm03 redder at
maximum and 0\fm12 in the tail of the color curve, as shown in Fig. 2.  
Obviously, if we try to derive a $B-V$ color excess from late-time
photometry, we must identify and correct systematic differences in the
photometry that are as large as shown in Fig. 2.

To account for possible systematic errors in our photometry of SN 2001el we
investigated the method of computing filter corrections described by
Stritzinger et al. (2002).  In the end these authors chose not to apply their
derived filter corrections, describing the whole exercise as ``somewhat
disappointing.''\footnote[8]{Some of their disappointment stemmed from trying
to correct YALO observations taken with a very wide, non-standard $R$ filter.
Our YALO observations were made with a much narrower, more standard $R$
filter.} By contrast, we were quite encouraged by similar efforts for SN
2001el.  In particular, our corrections to the $B$ and $V$ magnitudes not only
tighten up the light curves in those filters, but they solve the problem of the
$B-V$ colors being too red in the tail of the color curve.  Also, our $JHK$
light curves are improved.  Still, should the reader disagree with our
application of the filter corrections, we present the data in this paper in
multiple tables: two for uncorrected photometry, and two for corrections.

The determination of filter corrections involves at minimum the dot product
(i.e. multiplying together) of filter transmission functions, quantum
efficiency as a function of wavelength, and an atmospheric transmission
profile.  The zeropoints are first fixed with Vega (see Appendix A), then one
determines synthetic broad-band magnitudes using spectra of
spectrophotometric standards (Hamuy et al. 1992, 1994).  This allows one to
derive color terms based on the synthetic photometry.  In order for the
synthetic color terms to match the color terms derived from actual broad band
photometry within 0\fm01 mag$^{-1}$, we found, as did Stritzinger et al.
(2002), that we had to shift the optical filter profiles by as much as 100
\AA\ to match the synthetic color terms with the actual observed
ones.\footnote[9]{Shifting the filter profiles in this manner will strike
some readers as unsatisfactory.  However, the manufacturer of our $R$-band
filter gives a particular filter curve, but states that when the filter is
used in a liquid nitrogen filled dewar, the filter curve should be shifted
300 \AA\ to the blue.  We find that a shift of 270 \AA\ gives the correct
color term, or 30 \AA\ to the red of the expected filter curve.  Thus, the
use of these wavelength shifts is justified at some level.} In Appendix B we
give the filter shifts used for the optical photometry.

We took as our filter references the $BVRI$ profiles of Bessell (1990),
corrected for wavelength (i.e. divided by $\lambda$).  In Fig. 3 we show the
derived magnitude corrections for $B$-band photometry of SNe 1999ee and
2001el.  In Fig. 4 we show the analogous filter corrections for the $V$-band.  
These corrections were obtained using spectra of SN 1999ee from Hamuy et al.
(2002), while some spectra of SN 2001el are discussed by Wang et al.
(2002).\footnote[10]{The first two of these spectra range from 4179 to 8634
\AA, while the last three range from 3327 to 8634 \AA.  Because the $B$-band
starts at roughly 3600 \AA, we obtain three synthetic $B$-band magnitudes and
five synthetic $V$-band magnitudes of SN 2001el. Both are needed to derive
filter corrections for $B$ and $V$ data because of color terms that need to
be applied.} Six HST spectra were also kindly made available to us by P.
Nugent.  These range from 2947 to 10238 \AA, thus covering the $UBVRI$
passbands, and were taken from 29.5 $ \leq t \leq$ 65.3 d.

We find that for the CTIO 1.5-m telescope with the filters used by Landolt the
$B$-band filter corrections are essentially the same for SN 1999ee and 2001el.  
The situation is not too bad for the CTIO 0.9-m telescope, but for YALO the
$B$-band corrections are not as well behaved, indicating that the filter
corrections can be telescope {\em and} object dependent.  For example, features
at the blue end of the $B$-band are included more or less with different $B$
filters, and these features are different in the spectra SN 1999ee and 2001el.  
As a check on our method, filter corrections based on spectra by Wang et al.
(2002) and HST (Nugent, private communication) give very consistent corrections
at $t \approx$ 40 d for the YALO $B$-band filter.  In the case of the $V$-band
(see Fig. 4), the corrections are better behaved and seem to be telescope
dependent, i.e. more independent of the object under study.

In Table 4 we give the $B$-band and $V$-band corrections to the photometry
tabulated in Table 3.  These corrections would place the measurements
obtained with three different telescopes and filter sets on the Bessell
(1990) system.  The consistency of the photometry obtained on different
telescope is improved.  With the corrections, $\Delta B$ = +0\fm020 and
$\Delta V = -$0\fm002, with ``$\Delta$'' in the sense ``YALO {\em minus}
Landolt''.  As we show below, the late time YALO
$B-V$ colors give an estimate of the reddening which is in agreement with
other estimates of the reddening, but only if we take into account
the systematic differences of the filters.

The $R$-band photometry given in Table 3 (i.e. YALO vs. Landolt) is
consistent at the 0\fm01 mag level, on average.  Under the assumption that
the spectra of SNe 1999ee and 2001el are the same at $t$ = 16 d, the
application of $R$-band filter corrections would make the Landolt and YALO
$R$-band photometry differ by as much as 0\fm04.  We do not have
spectra of SN 2001el covering the $R$ and $I$-bands at this epoch, and the
photometry is not crying out for corrections to be made.

There is a systematic difference of $I$-band photometry in Table 3 at the
0\fm06 level. The application of filter corrections based on spectra of SN
1999ee would pull the data further apart by an additional 0\fm10 mag or more.
Thus, while the SN 2001el data in Table 3 motivates us to apply $I$-band filter
corrections, corrections based on spectra of a different object make the
systematic differences worse, not better.

In the case of the $R$- and $I$-bands, we also suspect some unquantified effect
of the YALO dichroic on the photometry.  While we are satisfied with the $B$-
and $V$-band corrections, we feel it unwise to apply corrections to the $R$-
and $I$-band data.

Fig. 5 shows the $UBVRI$ light curves of SN 2001el, with the corrections
of Table 4 applied to the $B$ and $V$ data given in Table 3.  

The most interesting result of the application of the filter corrections
is that by $t$ = 40 d we have reproduced a correction (with respect to
Bessell filter functions) of $\approx$0\fm1 to the YALO $B-V$ colors (see
Fig. 10 below), almost exactly what was found for SNe 1998bu and 1999ee
using YALO and CTIO 0.9-m photometry.

In Table 5 we give near-infrared photometry of SN 2001el obtained with
the YALO and LCO 1-m telescopes.  The near-infrared data were calibrated
using our ``best'' IR secondary standard, star 6, which was tied to
the system of Persson et al. (1998) from observations on five photometric
nights.\footnote[11]{The $K$-band magnitude of star 6 is based on
data from only two photometric nights. We note that the optical and 
infrared colors of this star are entirely consistent with it being
a normal K2 III star (Cox 2000). We do not suspect that it is variable.}
We note that the LCO observations and Persson's system of standards
rely on different $J$- and $K$-band filters than the broader-band  filters used
at YALO.  At LCO one uses a ``$J_{short}$'' and ``$K_{short}$'' filter.
These are the very filters and telescope used to establish the system of
Persson et al. (1998), and by definition the color terms are zero.

The YALO data are systematically fainter than the LCO 1-m data by roughly
0.1 mag at $t$ = 20 d in the $J$- and $H$-bands, and also by 0.1 mag or
more in $J$ beyond $t$ = 55 d, while there appear to be no significant
differences in the $K$-band data.  Can synthetic photometry confirm that
the YALO data are expected to be different at certain epochs for $J$ and
$H$, while the $K$-band data need little correction beyond $t$ = 20 d? To
test this, we constructed effective transmission profiles using the raw
filter profiles, an atmospheric transmission profile, quantum efficiency,
and multiple applications of reflection and window transmission in the
optical paths.  Then, using spectra of Vega, Sirius, and the Sun (see
Appendix A), we found that the synthetic infrared color terms for the YALO
telescope match observed ones to better than 0\fm01 mag$^{-1}$. This
required no arbitrary shifting of the filter transmission profiles.

In Fig. 6 we show the derived photometric corrections to YALO data, based
on spectra of SN 1999ee, which would have to be added to YALO data to
make them agree with photometry obtained with the LCO 1-m.  Under the
assumption that these corrections are applicable to SN 2001el, we give in
Table 6 a set of corrections to YALO IR photometry.  We assume that the
$J$- and $H$-band corrections at $t$ = 43 d are appropriate until $t$ =
64 d.  Obviously, the two assumptions just made may be somewhat dubious.  
But consider the IR light curves of SN 2001el shown in Fig. 7.  By
application of the corrections in Table 6, we find that the $J$- and
$H$-band data are in agreement within the internal errors now that
corrections up to 0.13 mag have been made. Since the corrections based on
spectra of an actual Type Ia SN are about the right size and are to be
made in the right direction, we feel confident that an application of
these corrections is fully warranted for the IR data. Thus, for the rest
of this paper we include these corrections, and all analysis based on IR
data uses the corrected YALO data.

One curious thing about the $H$-band light curve is that SN 2001el was
brighter at the time of the second maximum ($H$ = 12.97 at $t$ = +22.2 d) than
at the time of the first maximum ($H$ = 13.08 at $t = -$4.7 d).  This is in
agreement with the prediction of Krisciunas et al. (2000, Fig. 13; Table 11)
for Type Ia SNe with mid-range decline rates.  

Table 7 gives the maximum apparent $UBVRIJHK$ magnitudes and the
respective times of maximum for SN 2001el.  Note that within the
uncertainties the time of maximum light in the $I, J, H$, and $K$ bands was
the same, about 4.0 days before T($B_{max}$).  This is apparent visually
by an inspection of Fig. 8, which also shows how the secondary hump or
shoulder in the light curve varies as a function of filter from $V$ through
$K$.

To compare SN 2001el with other supernovae, we show in Fig. 9
a stack of $H$-band light curves ordered by the decline rate parameter
$\Delta$m$_{15}$(B).  Given the number of nights of SN 2001el photometry
and the minimal gaps in the dataset, it is clear that the data presented
here are of high quality.  Just as in the $I$-band, most Type Ia SNe
exhibit a secondary hump in $H$.  However, the objects with the lowest
values of $\Delta$m$_{15}$(B) seem to exhibit rather flat light curves in
the $H$-band for the first 40 days after T($B_{max}$).  Unlike $I$-band
light curves, however, Type Ia SNe with large $B$-band decline rates 
(e.g. SN 2000bk) do not monotonically decline after the time of maximum
light.

\section{Discussion} 

\subsection{Optical Light Curves}

The uncertainties given in Table 3 should be considered minimum
values of the internal errors (i.e. essentially based on photon
statistics and the uncertainties of the zero points and color
corrections).  A more honest measure of the internal errors can
be obtained from higher order polynomical fits to the light
curves, filter by filter, under the assumption that the supernova's
light varied smoothly with time.  On this basis we believe that
the uncertainties of the $U$-band magnitudes in Table 3 are sensible
as given.  For $B$, $V$, $R$, and $I$, respectively, we estimate that
the internal errors are $\pm$ 0.016, 0.025, 0.029, and 0.047 mag,
where the $B$ and $V$ data include the filter corrections of Table 4.

According to Phillips et al. (1999) a typical Type Ia SN has a
$B$-band decline rate $\Delta$m$_{15}(B)$ = 1.1 mag.  This parameter
varies roughly from 0.75 (SN 1999aa; Krisciunas et al. 2000)
to 1.94 mag (SN 1999da; Krisciunas et al. 2001).  SN 2001el has
$\Delta$m$_{15}(B)$ = 1.13 $\pm$ 0.04 mag and is thus
a mid-range decliner.

One of the characteristic features of the light curves of Type Ia SNe is
the secondary hump observed in the $IJHK$ bands.  To a lesser extent this
can be seen as a ``shoulder'' in the $R$-band, and sometimes even in the
$V$-band light curves. Krisciunas et al. (2001, Fig. 17) took various
well sampled $I$-band light curves, fitted polynomials to the data
between 20 and 40 days after T($B_{max}$), using the flux at $I$-band
maximum as the reference, and showed that an integration of the $I$-band
flux over this range correlated quite well with the $B$-band decline
rate.  If we let X = $\langle$ I $\rangle_{20-40}$, we find that

\begin{equation}
\Delta m_{15}(B) \; = \; 0.0035 \; \; + \; 17.475 X \; - \; 47.838 X^2 \; + \;
35.618 X^3 \; , 
\end{equation}

\parindent=0mm

with a typical {\sc rms} residual of $\pm$0.093 mag in $\Delta$m$_{15}$(B).

\parindent=9mm

Since the $B$-band decline rate is correlated with the intrinsic
luminosity at maximum of Type Ia SNe, therefore, this relationship
implies that the strength of the $I$-band secondary hump is also
correlated with the intrinsic luminosity.  We note two exceptions to Eqn.
8 above. SN 1992bc had a much weaker $I$-band secondary hump than
expected on the basis of its $B$-band decline rate, and SN 1994M had a
much stronger $I$-band secondary hump.

SN 2001el had a mean flux over the time span $20 < t < 40$ d
of 0.594 $\pm$ 0.04 with respect to the $I$-band maximum.  On the basis of
its B-band decline rate, we would have predicted $\langle$ I $\rangle
_{20-40}$ = 0.530.  Thus, it has a slightly stronger than ``normal''
$I$-band secondary hump, but well within the uncertainty of the
relationship given above.  

\subsection{Spectra}

Regarding the spectrum of SN 2001el, Sollerman et al. (2001) note
that the 612 nm Si II absorption line appeared flat-bottomed.  
Interstellar absorption lines of Ca H and K, and the D lines of Na I were
observed at the redshift of NGC 1448.  Wang et al. (2001) describe
spectropolarimetry of SN 2001el on 26 Sept 2001 UT (JD = 2,452,178.5; $t =
-4.0$ d), finding that SN 2001el had a normal spectrum similar to that of
SN 1994D except for a strong double-troughed absorption feature around
800.0 nm. The absorption dips of Si II (635.5, 564.0 nm) and Fe II (492.4,
516.9 nm) all showed velocities of about 10,000 km s$^{-1}$.  Ca II lines
(the ``infrared triplet'') were present in the spectrum at a velocity
comparable to that of the Si II 635.5-nm feature but were much weaker than
an absorption feature at 800.0 nm.  It is possible that the 800-nm feature
was a second component of high-velocity Ca II, as suggested by Hatano et
al. (1999) to explain a similar (but much weaker) feature in SN 1994D; if
this identification is correct, then this Ca II feature was due to a
detached shell/clump with a relative velocity of roughly 23,000 km
s$^{-1}$.

Wang et al. (2002) also obtained spectropolarimtery of SN 2001el on 1, 10, 18
October and 9 November 2001 UT.  Their data indicate, among other
things, that the value of R$_V \equiv$ A$_V$ / E($B-V$) = 2.88 $\pm$ 0.15,
somewhat less than the standard Galactic value of 3.10 (Sneden et al. 1978;
Rieke \& Lebofsky 1985).  From their spectra the interstellar Na I line 
has an equivalent width of 0.47 \AA.  From Fig. 4 of Barbon et al. (1990)
this corresponds to a color excess E($B-V$) $\approx$ 0\fm12.  However, the
uncertainty of this number is greater than or equal to 0\fm10.  We believe
that the most accurate estimate of E($B-V$) is obtained from a combination
of optical and infrared photometry (see below).

The HST spectra provided by Nugent for purposes of determining filter
corrections and comparing synthetic photometry to our data were obtained
on 29 October, 6, 13, 20, 26 November, and 4 December 2001 UT.  These
spectra will be discussed in a subsequent paper.

\subsection{Reddening and Implied Extinction}

From an analysis of SNe 1992A, 1992bc, 1992bo, and 1994D Lira (1995) found
a uniform color evolution for unreddened $B-V$ colors of Type Ia SNe from
30 to 90 days after $V$-band maximum.  Phillips et al. (1999) used Lira's
relation to get an estimate of the color excess for the tail of the $B-V$
colors of the SNe they analyzed. A majority of SNe analyzed by Jha (2002,
\S 4.2.3) show that the Lira relation works very well to describe the late
time $B-V$ color evolution of unreddened Type Ia SNe. In Fig. 10 we show
the $B-V$ colors of SN 2001el, with and without the filter corrections
given in Table 4.  Given that the Lira relation was based on observations
with the CTIO 0.9-m telescope, we clearly want to correct the YALO data to
the CTIO 0.9-m system so as to derive E($B-V$)$_{tail}$ (see below) using
the same zeropoint for the Lira relation.  

In Table 8 we give the color excesses for SN 2001el based on a variety of
indices.  Following the analysis of Phillips et al. (1999), we give the $B-V$
color excess as implied from the photometry at maximum light, the $V-I$ color
excess at maximum, and the $B-V$ color excess from the photometry for $t > $
32 d.  The three estimates of A$_V$ that result are in reasonable agreement,
but only if we correct the $B-V$ colors for filter differences.  A weighted
$B-V$ color excess is E($B-V$)$_{avg}$ = 0.206 $\pm$ 0.046 mag for the host
galaxy reddening. The small contribution of Galactic dust reddening is
E($B-V$) = 0.014 mag (Schlegel et al. 1998), indicating a small Galactic
contribution to A$_V$ amounting to 0.043 mag. Adopting R$_V$ = 2.88 $\pm$ 0.15
from Wang et al. (2002) for the host galaxy, the estimated total extinction
based on optical photometry alone is A$_V$ = 0.64 $\pm$ 0.14 mag.

As is well known, interstellar extinction in the near-infrared is an
order of magnitude lower than in the $V$-band.  If there exists a color
index such as $V-K$ which is ``well behaved'' for some range of decline
rates of Type Ia SNe, then a $V-K$ color excess, for example, is almost a
direct measure of the $V$-band extinction.  We shall adopt Rieke \&
Lebofsky's (1985) values of A$_{\lambda}$ / A$_V$ = 0.282, 0.175, and
0.112 for the $J$-, $H$-, and $K$-bands, respectively, and assign an
uncertainty of 20 percent to each ratio.  It follows that A$_V$ = (1.393
$\pm$ 0.110) E($V-J$) = (1.212 $\pm$ 0.052) E($V-H$) = (1.126 $\pm$
0.028) E($V-K$).  In other words, if we can estimate a color excess in
$V-K$, A$_V$ numerically is only 10 to 15 percent greater than that color excess.

Krisciunas et al. (2000, 2001) showed that over a range of decline rates
(0.87 $< \Delta$m$_{15}$(B) $<$ 1.28) Type Ia SNe exhibit uniform $V$ {\em
minus} near-IR color curves from $ -9 \leq t \leq +27$ d. $V-K$ colors were
particularly well behaved, and over the widest range of decline rates.
However, on the basis of a small number of data points {\em per object}
they could not justify fitting the $V-H$ or $V-K$ colors by anything more
sophisticated than two straight lines.  The $V-J$ colors were fitted with a
second order curve for $-9 \leq t \leq +9.5$ d, while a linear fit was used
for $+9.5 \leq t \leq +27$ d.

From the well sampled $V$ {\em minus} near-IR color curves of SNe 1999ac and
2001el we now know that there is no {\em abrupt} change of the color
evolution one week after T($B_{max}$).  Thus we feel justified in
fitting a higher order curve to the data of eight mid-range decliners
studied by Krisciunas et al. (2000) to give more realistic unreddened color
loci.

In Fig. 11 we show the V {\em minus} near-IR colors of SN 2001el, along
with loci derived from SNe 1972E, 1980N, 1983R, 1999cp, 1981B, 1981D,
1998bu, and 1999cl, adjusted in the ordinate direction so as to minimize
the reduced $\chi^2$ of the fits.  The shape of the $V-K$ locus matches the
data of SN 2001el very well (with $\chi^2_{\nu}$ = 0.76).  
The $V-H$ and $V-J$ loci match the data of SN 2001el well only 
from $+10 \lesssim t \lesssim +20$ d.

In Table 8 we also give the color excesses and values of A$_V$ derived
from $VJHK$ photometry.  Within the errors these estimates of A$_V$ are in
agreement with the value derived from optical photometry alone.  The
weighted mean from the $V$ {\em minus} near-IR colors and E($B-V$)$_{avg}$
is A$_V$ = 0.57 $\pm$ 0.05 mag.\footnote[12]{The various estimates of
A$_V$ are only semi-independent, as all rely on the $V$-band photometry.
However, we believe that our $V$-band photometry suffers from no
serious systematic errors.} The use of optical and infrared data
results in a smaller statistical uncertainty for the extinction.  Though
the filter corrections adopted for the $BVJHK$ photometry are only
approximate ($\pm$0.02 mag), we feel that the consistency of the various
estimates of A$_V$ is facilitated by the adoption of the filter corrections.
While the calculation of such corrections is time consuming because of
its iterative nature, this extra work will likely become a standard
part of the production of high quality SN light curves.

\subsection{Theoretical V {\em minus} IR Color Curves}

Model calculations for the explosion, light curves and spectra of typical
Type Ia SNe indicate that there exists a physical basis for well behaved
optical versus infrared color relationships (H\"{o}flich 1995, Wheeler et
al. 1998). What we see as a supernova event is the light emitted from a
rapidly expanding envelope. Due to the increasing geometrical dilution,
deeper layers of the envelope become exposed with time. Until about maximum
light, the spectra are formed in the regions which have undergone
incomplete Si-burning.  During this period both the fluxes in the $V$- and
$K$-bands are mainly determined by the free-free/bound-free opacities and
Thomson scattering and the source functions, i.e. the temperature evolution
at the photosphere which is rather insensitive to the total luminosity
because of the steep dependence of the opacity on the temperature (H\"{o}flich
et al. 1993).

 
Starting a few days before maximum light, the $V$-band flux reflects the
evolution of the luminosity. Around the same time, the Fe/Ni core becomes
exposed, and broad emission features due to iron group elements occur in the
$H-$ and $K$-bands (Wheeler et al. 1998) which determine the flux. They are in
emission because the IR source functions are close to thermal and the optical
depth is large, i.e. they are formed well above the photosphere close to the
outer edge of the Fe/Ni core. With time, the optical photosphere receeds in
mass. Thus, the emission features in the $K$-band increase in strength while the
$V$-band flux decreases, leading to $V-K$ colors that get redder for some time.
Once the SNe enter the nebular stage there should be a spread in $V-K$ color
curves for different objects, even if they have the same decline rate.  The
$V-K$ color curve should be reasonably model-independent (i.e. whether the
explosion is due to deflagration or a delayed detonation should not matter
significantly) because, virtually, the entire $K$-band is dominated by the iron
feature.
 
The $V-J$ and $V-H$ color curves would not be as well behaved because the $J$-
and $H$-bands cover spectral regions which are affected both by the IR-emission
features and the continua. The fractional contribution depends on the Doppler
shift of the chemical layers and, thus, it depends on the details of the nuclear
burning. Note that for very subluminous Type Ia SNe (e.g. SN1999by), the $V-K$
evolution after maximum light is delayed because the Fe/Ni core is explosed only
two weeks after maximum (H\"{o}flich et al. 2002).

In Fig. 12 we show the $V-H$ and $V-K$ color curves for a delayed detonation
model of a ``Branch-normal'' Type Ia SN (model 5p028z22.23 of H\"{o}flich et
al. 2002). To derive synthetic magnitudes we used the $V$-band filter profile
of Bessell (1990) and the $H$ and $K_{short}$ filter profiles of Persson et
al. (1998), along with appropriate quantum efficiency and atmospheric
transmission functions.  The fluxes were normalized to the spectra of Vega
given by Castelli \& Kurucz (1994) and assumed $V-H$ and $V-K$ = 0.00. For
comparison we show the unreddened loci based on the 8 SNe studied by
Krisciunas et al. (2000). The color indices are well reproduced within the
model uncertainties of $\approx$ 0\fm2 to 0\fm3.  We note that our delayed
detonation model gives $t$(V$_{max}$) $- t$(B$_{max}$) = 1.1 d, which is
smaller than the nominal value of 1.7 to 2.0 d.  A shift of the theoretical
$V$ {\em minus} infrared color curves by $\pm$1 d is within the uncertainties.
Likewise, from data on any given SN, the time of $B$-band maximum is rarely
known to better than $\pm$0.5 d, so the observed colors can be shifted in
the time direction as well.

The colors have been calculated by our spherical NLTE light curve code which
uses about 500 depths, 3000 frequency points and 50 NLTE-superlevels. A
comparison with detailed NLTE-spectral calculations shows that the reduction
of atomic levels and frequencies produces uncertainties of $\approx$ 0\fm2 to
0\fm3 in $H$ and $K$. However, the small number of frequencies is responsible
for the failure to properly reproduce $V-J$, especially after the time of
$B$-band maximum when the flux varies strongly within the $J$ band (a factor
of $\approx$ 10; Wheeler et al. 1998).  As a result, we do not show our
theoretical $V-J$ curve.

\subsection{Distance}

NGC 1448, the host of SN 2001el, has not been
the target of study for determining its distance via surface
brightness fluctuations (SBF) or Cepheids.  Since it is a spiral
galaxy inclined to the line of sight, one can derive a Tully-Fisher
(TF) distance to it.  Mathewson \& Ford (1996)
give the following parameters: velocity width V$_{20}$ = 208 km s$^{-1}$,
inclination angle $i = 88^o$, integrated extinction-corrected I-band magnitude
9.44.  Using the $I$-band Tully-Fisher calibration of Sakai et al. (2000),
W$_{20}$ = 2 V$_{20}$, and a slight correction for the inclination angle
of the galaxy and its redshift, we find W$^{C}_{20}$ = 414.6 km s$^{-1}$.
It follows that the absolute I-band magnitude of NGC 1448 M$_I = -$22.21 and the
distance modulus is 31.65 mag.  Giovanelli et al. (1997) indicate that the
uncertainty of a single TF distance is roughly $\pm$ 0.35 mag.
The TF distance of NGC 1448 is therefore 21.4$^{+3.7}_{-4.2}$ Mpc.

We note that $H$-band and $K$-band light curves of many Type Ia SNe are
reasonably flat at $t \approx$ +10 d.  We assume, then, that this is a
``well behaved'' part of the light curves. The $J$-band light curves, by
contrast, exhibit rapidly changing flux at this epoch.  

In Fig. 13 we show the absolute magnitudes at $BVI$ maximum light and
absolute $H$-band magnitudes at $t$ = +10 d versus the decline rate
parameter $\Delta$m$_{15}$(B). Some of the data shown are derived using
Cepheid distances to host galaxies, surface brightness fluctuations, the
planetary nebula luminosity function method, or the ``tip of the red giant
branch'' method, assuming the Cepheid distance scale adopted by Freedman et
al. (2001).  Other data for galaxies in the Hubble flow (i.e. with
redshifts greater than 3000 km s$^{-1}$) are derived assuming H$_0$ = 74 km
s$^{-1}$ Mpc$^{-1}$. In all four bands the absolute magnitude is reasonably
constant for 0.87 $<$ $\Delta$m$_{15}$(B) $<$ 1.30.  For 9 SNe which
satisfy this criterion we find a weighted mean $H$-band absolute magnitude
at $t$ = +10 d of M$_{H}$($t$=10) = $-$17.91 $\pm$ 0.05.  

From an interpolation of the {\em corrected} YALO data for SN 2001el we
estimate that at $t$ = +10 d $H$ = 13.228 $\pm$ 0.04.  Adopting an
extinction of A$_H$ = 0.10 mag and the absolute magnitude just given, we
obtain a distance modulus of 31.04 mag with an uncertainty of $\pm$
0.14.  The corresponding distance is 16.1 $\pm$ 1.1 Mpc.

Using the method of Phillips et al. (1999), the $BVI$ optical photometry
gives a distance modulus of m$-$M = 31.29 $\pm$ 0.08 mag.  This is on the
Cepheid distance scale adopted by Freedman et al. (2001). The
corresponding distance is 17.9 $\pm$ 0.8 Mpc.  

With a distance somewhere in the range 16.1 to 21.4 Mpc, NGC 1448
is clearly near enough for a determination of its distance via Cepheids.

\section{Conclusions} 

Perhaps the most significant finding of this paper is that we were able to
calculate corrections to $BV$ and infrared photometry, based on synthetic
photometry, but using spectra of actual supernovae, that to some extent
eliminated systematic differences in the data obtained on different telescopes
with different physical detectors and filters.  Under the assumption that the
infrared spectrum of SN 2001el was like that of SN 1999ee, and that it evolved
similarly with time, we derived corrections to YALO IR photometry which had the
right size and were to be applied in the right direction to bring the YALO data
fully onto the system of Persson et al. (1998), and in agreement with the data
taken with the LCO 1-m telescope. Though it is known that the spectrum of a SN
is not like that of a star, especially once the SN enters the nebular phase ($t
\gtrsim$ 30 d), such corrections have been ignored in the past because we
simply did not have enough spectra and photometry to confirm to what degree
synthetic photometry gives the right results.  Clearly, not all optical
response functions are well-known. Further information (e.g. the effect of the
YALO dichroic on the optical transmission) is needed to make analogous
corrections to the $RI$ data.

Another significant result of this paper is the excellent agreement between
theoretical $V-H$ and $V-K$ color curves with the unreddened loci based
on 8 SNe studied by Krisciunas et al. (2000).  This gives us further
confidence that there exist ``uniform'' color curves for Type Ia SNe
with mid-range decline rates, which can be used to derive accurate values
of the total extinction suffered by the light of such objects.  A comparison
of theoretical color curves with observational data, covering the full
range of decline rates of Type Ia SNe, is beyond the scope of the
present paper.

Given the importance of SNe in modern cosmological studies, it is worth
our while to investigate the uniformity of behavior of as many nearby SNe
as we can.  Though the light curves of SN 2001el are quite normal,
we note that the {\em observed} $B-V$ color 3 to 4 days before T($B_{max}$)
was slightly negative.  Given the extinction along the line of sight,
the color excess E($B-V$) must be about 0\fm20.  Thus, the dereddened
color shortly before T($B_{max}$) was $B-V \approx -0.22$, which is extremely
blue. The second H-band maximum at $t$ = +22.2 d was 0.1 mag brighter than
the first H-band maximum at $t = -$4.7 d.  This is consistent with a
prediction of Krisciunas et al. (2000) based on an analysis of other Type
Ia SNe with mid-range decline rates, but not all $H$-band light curves
of mid-range decliners show this.

On the basis of the optical and IR photometry we derive A$_V$ = 0.57 $\pm$
0.05 mag.  Given the small formal uncertainty in A$_V$, we can state that
the uncertainty in the distance estimate to SN 2001el and its host is {\em not}
due to the extinction along the line of sight.  Given the attention we
have paid to possible sources of systematic error in the $UBVRIJHK$
photometry and the good coverage of the light curve, we hope that 
photometry of SN 2001el can serve as a template for studies of other 
Type Ia SNe.

With a distance in the range 16.1 to 21.4 Mpc, the host galaxy of SN
2001el is close enough for its distance to be determined by means of
Cepheids.  Given the small number of nearby galaxies which have hosted SNe
and also had their distances determined via Cepheids, we suggest that NGC
1448 be a future target for such observations with HST.

It is worth (re-)stating that (for the forseeable future) high redshift SNe will
only be observed in {\em rest frame} optical bands, because we do not yet have
infrared detectors sensitive enough, even with an 8-m class telescope, to detect
these objects in the rest frame IR bands.  This is unfortunate, as it is becoming
increasingly clear that a combination of optical and infrared photometry of Type
Ia SNe allows us to get a better handle on the host extinction suffered by these
objects.  Since the advocacy of a non-zero cosmological constant based on
supernova studies hinges on the systematic faintness of objects in the redshift
range $0.2 < z < 0.8$, we want to minimize as well as possible the effect of
systematic errors in distances to these most important cosmological beacons.

\vspace {1 cm}

\acknowledgments

We thank STScI for the following support: HST GO-07505.02-A, HST-GO-08641.07-A,
HST-GO-8177.06 (the High-Z Supernova Team survey) and HST-GO-08648.10-A (the
SInS collaboration).  We thank Peter Nugent for providing HST spectra in
advance of publication;  support for HST proposal \#8611 was provided by NASA
through a grant from the Space Telescope Science Institute, which is operated by
the Association of Universities for Research in Astronomy, Inc., under NASA
contract NAS5-26555.

Eric Persson provided much necessary information for the
calculation of the infrared filter corrections.  We also thank Mario Hamuy for
useful discussions;  Lifan Wang for sharing data ahead of publication; and
Michael Merrill for his telluric atlas.  AUL acknowledges support from NSF grant
AST0097895.  Useful information was obtained from the NASA/IPAC Extragalactic
Database.

\appendix

\section{Spectra for Synthetic Photomery}

The near-infrared photometry required an estimate of the theoretical
color terms for the various natural systems. There is no grid of
spectrophotometric standards in the near-infrared comparable to what
is available in the optical. As a simple guide to color terms, we have
chosen to use synthetic photometry based on the Sun, Vega, and
Sirius. This color range generally encompasses the photometric
standards (Persson et al. 1998) used in our study. All
spectrophotometric data were convolved 2
\AA\ and sampled it at 1 \AA\ per pixel. The spectrophotometric
models did not include atmospheric absorption. Rather, we added the
telluric absorption to the tranmissions functions used to calculate
the synthetic photometry. The adopted photometry was taken from the
discussion in Bessell et al. (1998).

\subsection{The Sun}

The solar spectrophotometry is a combination of observed solar fluxes
and models. We chose the solar model from the Kurucz website (see
also Fontenla et al. 1999) as ($T_{eff}$, log g, $v_{micro}$, mixing
length/scale height) = (5777 K, 4.438, 1.5 km s$^{-1}$, 1.25). The
Kurucz model reproduces the Kitt Peak Solar Flux Atlas (Kurucz et
al. 1984). In the near-infrared, we have inserted the Livingston \&
Wallace (1991) observed spectra, which were scaled to the continuum of
the Kurucz model.

\subsection{Vega}

The Vega spectrophotometry is a combination of empirical data and
model data. The fundamental flux calibration in the optical comes from
Hayes (1970, 1985), Oke \& Schild (1970), and T\"{u}g (1980). We have
adopted the model ``veg090250000p.asc5'' from the website of
Kurucz. This model corresponds to the physical parameters ($T_{eff}$,
log g, [M/H], $v_{micro}$, $v_{macro}$) = (9550 K, 3.950, $-$0.5, 2 km
s$^{-1}$, 0 km s$^{-1}$) which is the model adopted by Bessell et
al. (1998).  This model was then scaled from the calculated Eddington
fluxes corrected to an angular size $3.24 \pm 0.07$ mas (Code et
al. 1976).

We then compared this scaled model to the Hayes (1985) fluxes of Vega
in the wavelength region 4000-8000 \AA\ and found that the model
needed to be scaled to 3.26 mas, in agreement with the conclusions of
Bessell et al. (1998). At the extremes of the measured data the models
did not fit well. We needed to scale the model by 0.972 at 3300 \AA\
and 1.009 at 10500 \AA\ to fit the Hayes points.

The final semi-empirical spectrophotometric model of Vega was created
by using the Hayes (1985) points from 3300-10500 \AA, the Kurucz model
scaled by 0.972 below 3300 \AA, and the Kurucz model scaled by 1.009
above 10500 \AA. A more complete discussion of the range in models of
Vega can be found in Cohen et al. (1999) and Bessell et al. (1998).

\subsection{Sirius}

For Sirius we have adopted the model``sir.ascsq5'' from the Kurucz 
website which was run with the physical parameters ($T_{eff}$, log g,
[M/H], $v_{micro}$) = (9850 K, 4.30, +0.4, 0 km s$^{-1}$). This is
very similar to the model used by Cohen et al. (1999). We scaled the
model from the Eddington fluxes to observed fluxes assuming an angular
diameter of 5.89 mas (Code et al. 1976).  It was found, however, that
this did not reproduce the observed $V$ mag of Sirius from Ian Glass
as quoted in Bessell et al. (1998). The final fluxes were scaled by
1.049 to force $V=-1.43$.

\section{Filter Shifts for Synthetic Photometry}

Synthetic photometry uses spectra of spectrophotometric standards, filter
and atmospheric transmission curves to derive broad band magnitudes of the
standards.  These should give photometric color terms that match the color
terms derived from actual photometry of stars.  However, we find that the
filter profiles must be shifted in order to match the synthetic color terms
with the observed values.  In Table 9 we give the filter shifts adopted
to
place optical photometry on the Bessell (1990) system. 

We found that no shifts needed to be applied for the infrared photometry (YALO
versus LCO 1-m).

We note that the $R$-band filter used by Stritzinger et al. (2002) for their
YALO observations was a very wide non-standard filter.  Our observations
were made with a much narrower, more standard filter.

We also note that Stritzinger et al. (2002) found a filter shift of only 10
\AA\ for the YALO $I$-band filter, compared to our value of 100 \AA.  Their
color terms for YALO photometry were based on the photometric sequence of SN
1999ee, rather than Landolt standards, as were ours, and their $I$-band color
term has a size half as large as ours.  Our filter shift would be comparable
to theirs for identical color terms.  

As stated above, since the YALO and Landolt $R$-band data agree, on average, at
the 0\fm01 level, there is no great motivation to apply $R$-band filter
corrections.  If we did so, it would produce systematic
differences at the 0\fm04 level.  We do have a motivation to resolve
systematic differences in $I$-band photometry, but we do not have spectra of
SN 2001el at the appropriate epochs to make these corrections directly.  
Assuming that the spectra of SNe 1999ee and 2001el are the same in the
$I$-band would give us filter corrections that would make the photometry
nearly 0\fm20 different (YALO vs. Landolt) at $t$ = 16 d.  We suspect that a
significant unknown is the effect of the YALO dichroic on the actual $R$-
and $I$-band filter functions.

\begin{deluxetable}{ccccccc}
\tablewidth{0pc}
\tablecaption{Optical Photometric Sequence near SN 2001el}
\tablehead{   \colhead{Star ID} &
\colhead{$\alpha$ (2000)} & \colhead{$\delta$ (2000)} &
\colhead {$V$} & \colhead{$B-V$} &
\colhead{$V-R$} & \colhead{$V-I$} } 
\startdata
SN    & 3:44:30.6 & $-$44:38:24 &      ...       &      ...       &      ...       &       ...      \\ 
1$^a$ & 3:44:37.7 & $-$44:39:34 & 12.736 (0.002) &  0.621 (0.003) &  0.372 (0.002) &  0.747 (0.004) \\   
1$^b$ &   ...     &    ...      & 12.739 (0.003) &  0.623 (0.006) &  0.367 (0.002) &  0.738 (0.002) \\
2$^a$ & 3:44:30.4 & $-$44:40:36 & 15.377 (0.002) &  1.086 (0.009) &  0.666 (0.003) &  1.221 (0.005) \\
5$^a$ & 3:44:21.8 & $-$44:38:38 & 15.913 (0.002) &  0.965 (0.010) &  0.532 (0.004) &  0.986 (0.005) \\
6$^a$ & 3:44:26.3 & $-$44:38:40 & 15.756 (0.002) &  1.132 (0.013) &  0.675 (0.006) &  1.236 (0.004) \\
7$^a$ & 3:44:29.5 & $-$44:37:29 & 14.524 (0.002) &  0.836 (0.005) &  0.487 (0.002) &  0.932 (0.005) \\
\enddata
\tablenotetext{a} {Based on 5 nights of CCD photometry with YALO.
$^b$The second set of values is based on 7 nights of single-channel
photoelectric photometry by Landolt, who also obtained $U-B$ = 
+0.049 $\pm$ 0.012 for this star.}
\end{deluxetable}

\begin{deluxetable}{cccc}
\tablewidth{0pc}
\tablecaption{Infrared Photometric Sequence near SN 2001el$^a$}
\tablehead{   \colhead{Star ID$^b$} &
\colhead{$J$} & \colhead{$H$} & \colhead{$K$} } 
\startdata
5    & 14.260 (0.042) & 13.739 (0.029) & 13.727 (0.021) \\
6    & 13.745 (0.009) & 13.163 (0.004) & 13.017 (0.026) \\
7    & 13.302 (0.043) & 13.170 (0.146) & 13.194 (0.041) \\
\enddata
\tablenotetext{a} {Based on 5 nights of $J$-band calibration with
the LCO 1-m telescope, 4 nights in $H$, and 2 nights in $K$.
$^b$The identifications are  the same as those in Table 1 and Fig. 1.}
\end{deluxetable}

\begin{deluxetable}{cccccccl}
\tabletypesize{\footnotesize}
\rotate
\tablewidth{0pc}
\tablecaption{UBVRI Photometry of SN 2001el}
\tablehead{   \colhead{JD$-$2,450,000} &
\colhead{UT Date$^a$} & \colhead {$U$} & \colhead {$B$} & \colhead{$V$} &
\colhead{$R$} & \colhead{$I$} &
\colhead{Observer+Telescope} }
\startdata

2171.85 & Sep19.35 & & 14.075 (0.013) & 13.808 (0.009) & 13.711 (0.016) & 13.742 (0.015) & Espinoza+YALO \\
2174.81 & Sep22.31 & & 13.407 (0.012) & 13.258 (0.008) & 13.111 (0.015) & 13.120 (0.014) & Espinoza+YALO   \\
2177.84 & Sep25.34 & & 13.032 (0.015) & 12.927 (0.010) & 12.768 (0.024) & 12.838 (0.016) & Espinoza+YALO   \\

2178.82 & Sep26.32 & 12.753 (0.045) & 12.932 (0.023) & 12.899 (0.014) & & & Arenas+CTIO 0.9-m \\   
2179.81 & Sep27.31 & 12.708 (0.043) & 12.887 (0.026) & 12.868 (0.015) & & & Arenas+CTIO 0.9-m \\

2183.84 & Oct01.34 & & 12.871 (0.019) & 12.686 (0.010) & 12.622 (0.045) & 12.885 (0.016) & D. Gonzalez+YALO  \\
2183.86 & Oct01.36 & 12.765 (0.027) & 12.850 (0.013) & 12.753 (0.008) & & & Arenas+CTIO 0.9-m \\
2184.88 & Oct02.38 & 12.824 (0.029) & 12.867 (0.013) & 12.755 (0.007) & & & Arenas+CTIO 0.9-m \\
2186.80 & Oct04.30 & & 12.966 (0.009) & 12.724 (0.004) & 12.624 (0.009) & 12.954 (0.008) & Espinoza+YALO  \\

2189.74 & Oct07.24 & 13.418 (0.019) & 13.157 (0.014) & 12.810 (0.013) & 12.729 (0.015) & 13.115 (0.015) & Landolt+CTIO 1.5-m \\
2189.76 & Oct07.26 & 13.392 (0.019) & 13.160 (0.014) & 12.812 (0.013) & 12.730 (0.015) & 13.119 (0.015) & Landolt+CTIO 1.5-m \\
2189.83 & Oct07.33 &                & 13.193 (0.007) & 12.784 (0.004) & 12.738 (0.010) & 13.073 (0.008) & Espinoza+YALO \\  
2189.85 & Oct07.35 & 13.428 (0.019) & 13.171 (0.014) & 12.831 (0.013) & 12.736 (0.015) & 13.113 (0.015) & Landolt+CTIO 1.5-m \\

2190.72 & Oct08.22 & 13.507 (0.019) & 13.230 (0.014) & 12.834 (0.013) & 12.772 (0.015) & 13.211 (0.015) & Landolt+CTIO 1.5-m \\
2190.85 & Oct08.35 & 13.545 (0.019) & 13.266 (0.014) & 12.859 (0.013) & 12.798 (0.015) & 13.207 (0.015) & Landolt+CTIO 1.5-m \\
2191.72 & Oct09.22 & 13.643 (0.019) & 13.331 (0.014) & 12.879 (0.013) & 12.835 (0.015) & 13.240 (0.015) & Landolt+CTIO 1.5-m \\
2191.85 & Oct09.35 & 13.669 (0.019) & 13.353 (0.014) & 12.901 (0.013) & 12.864 (0.015) & 13.278 (0.015) & Landolt+CTIO 1.5-m \\

2192.83 & Oct10.33 &                & 13.453 (0.016) & 12.906 (0.011) & 12.909 (0.018) & 13.259 (0.016) & D. Gonzalez+YALO \\
  
2192.85 & Oct10.35 & 13.765 (0.019) & 13.425 (0.014) & 12.924 (0.013) & 12.902 (0.015) & 13.331 (0.015) & Landolt+CTIO 1.5-m \\
2193.72 & Oct11.22 & 13.881 (0.019) & 13.542 (0.014) & 12.981 (0.013) & 12.969 (0.015) & 13.368 (0.015) & Landolt+CTIO 1.5-m \\
2193.85 & Oct11.35 & 13.913 (0.019) & 13.565 (0.014) & 13.007 (0.013) & 12.988 (0.015) & 13.374 (0.015) & Landolt+CTIO 1.5-m \\
2194.78 & Oct12.28 & 14.009 (0.019) & 13.651 (0.014) & 13.056 (0.013) & 13.032 (0.015) & 13.375 (0.015) & Landolt+CTIO 1.5-m \\

2195.75 & Oct13.25 &                & 13.793 (0.027) & 13.068 (0.018) & 13.040 (0.035) & 13.358 (0.027) & D. Gonzalez+YALO \\
2195.83 & Oct13.33 & 14.188 (0.019) & 13.779 (0.014) & 13.130 (0.013) & 13.103 (0.015) & 13.440 (0.015) & Landolt+CTIO 1.5-m \\
2197.71 & Oct15.21 & 14.450 (0.019) & 13.989 (0.014) & 13.230 (0.013) & 13.175 (0.015) & 13.462 (0.015) & Landolt+CTIO 1.5-m \\
2198.71 & Oct16.21 & 14.544 (0.019) & 14.086 (0.014) & 13.265 (0.013) & 13.156 (0.015) & 13.359 (0.015) & Landolt+CTIO 1.5-m \\

    
2198.80 & Oct16.30 &                & 14.155 (0.008) & 13.278 (0.005) & 13.169 (0.014) & 13.322 (0.008) & D. Gonzalez+YALO \\
2201.81 & Oct19.31 &                & 14.506 (0.012) & 13.453 (0.008) & 13.232 (0.014) & 13.253 (0.012) & Pizarro+YALO \\ 

2204.78 & Oct22.28 & & 14.804 (0.016) & 13.590 (0.009) & 13.303 (0.019) & 13.226 (0.024) & Espinoza+YALO \\  
2207.77 & Oct25.27 & & 15.090 (0.012) & 13.746 (0.007) & 13.392 (0.015) & 13.224 (0.012) & Pizarro+YALO   \\
2210.82 & Oct28.32 & & 15.357 (0.013) & 13.924 (0.006) & 13.524 (0.011) & 13.235 (0.010) & Pizarro+YALO   \\
2213.68 & Oct31.18 & & 15.554 (0.015) & 14.109 (0.006) & 13.697 (0.015) & 13.306 (0.009) & Espinoza+YALO   \\
2216.73 & Nov03.23 & & 15.727 (0.011) & 14.301 (0.005) & 13.880 (0.012) & 13.486 (0.009) & Espinoza+YALO   \\

2222.75 & Nov09.25 & & 15.998 (0.020) & 14.606 (0.010) & 14.248 (0.018) & 13.904 (0.017) & D. Gonzalez+YALO   \\
2225.72 & Nov12.22 & & 16.060 (0.012) & 14.706 (0.007) & 14.331 (0.014) & 14.038 (0.012) & D. Gonzalez+YALO   \\
2228.73 & Nov15.23 & & 16.149 (0.016) & 14.806 (0.009) & 14.478 (0.016) & 14.171 (0.015) & Espinoza+YALO    \\
2231.71 & Nov18.21 & & 16.198 (0.017) & 14.865 (0.009) & 14.584 (0.016) & 14.314 (0.014) & Espinoza+YALO    \\
2234.73 & Nov21.23 & & 16.215 (0.016) & 14.940 (0.006) & 14.675 (0.013) & 14.447 (0.012) & D. Gonzalez+YALO   \\

2237.73 & Nov24.23 & & 16.265 (0.017) & 15.042 (0.008) & 14.738 (0.016) & 14.560 (0.014) & D. Gonzalez+YALO  \\ 
2240.73 & Nov27.22 & & 16.282 (0.020) & 15.121 (0.009) & 14.835 (0.019) & 14.689 (0.015) & D. Gonzalez+YALO   \\
2243.71 & Nov30.21 & & 16.333 (0.021) & 15.178 (0.012) & 14.917 (0.021) & 14.801 (0.019) & Espinoza+YALO   \\
2246.78 & Dec03.28 & & 16.383 (0.038) & 15.238 (0.009) & 15.042 (0.020) & 14.922 (0.022) & Espinoza+YALO   \\
2289.64 & Jan15.14 & 17.695 (0.037) & 16.924 (0.013) & 16.349 (0.006) & & & Candia+CTIO 0.9-m \\

2324.55 & Feb19.05 & 18.685 (0.023) & 17.500 (0.011) & 17.072 (0.013) & 17.294 (0.008) & 17.401 (0.031) & Candia+CTIO 0.9-m \\
\enddata
\tablenotetext{a} {Year is 2001, except for the last two lines, for which year is 2002.}
\end{deluxetable}

\begin{deluxetable}{ccccc}
\tablewidth{0pc}
\tablecaption{Corrections to B-band and V-band Photometry$^a$}
\tablehead{   \colhead{JD$-$2,450,000} & \colhead{$t$ (d)} & 
\colhead{$\Delta B$} &
\colhead{$\Delta V$} & \colhead{Telescope$^b$} }
\startdata
2171.85 & $-$10.65 & [$-$0.027] & $-$0.007 & 1 \\
2174.81 & $-$7.69  & [$-$0.027] & $-$0.001 & 1  \\
2177.84 & $-$4.66  & [$-$0.027] & +0.005 & 1 \\
2178.82 & $-$3.68  & $-$0.038 &   +0.004 & 2 \\
2179.81 & $-$2.69  & $-$0.038 &   +0.005 & 2 \\

2183.84 & 1.34 & [$-$0.027] & +0.018 & 1 \\
2183.86 & 1.36 & $-$0.036 &   +0.005 & 2 \\
2184.88 & 2.38 & $-$0.035 &   +0.004 & 2 \\
2186.80 & 4.30 & [$-$0.027] & +0.020 & 1 \\
2189.78 & 7.28 & $-$0.012 & +0.001 & 3 \\

2189.83 & 7.33 & [$-$0.027] & +0.023 & 1 \\
2190.79 & 8.29 & $-$0.011 &  0.000 & 3 \\
2191.79 & 9.29 & $-$0.010 & $-$0.001 & 3 \\
2192.72 & 10.22 & $-$0.008 &  0.000 & 3 \\
2192.83 & 10.33 & $-$0.026 & +0.026 & 1 \\

2193.78 & 11.28 & $-$0.007 & +0.001 & 3 \\
2194.78 & 12.28 & $-$0.006 & +0.002 & 3 \\
2195.75 & 13.25 & $-$0.024 & +0.028 & 1 \\
2195.84 & 13.34 & $-$0.004 & +0.004 & 3 \\
2197.71 & 15.21 & $-$0.002 & +0.005 & 3 \\

2198.71 & 16.21 & 0.000  & +0.007 & 3 \\
2198.80 & 16.30 & $-$0.022 & +0.030 & 1 \\
2201.81 & 19.31 & $-$0.023 & +0.033 & 1 \\
2204.78 & 22.28 & $-$0.039 & +0.034 & 1 \\
2207.77 & 25.27 & $-$0.051 & +0.035 & 1 \\

2210.82 & 28.32 & $-$0.065 & +0.037 & 1 \\
2213.68 & 31.18 & $-$0.069 & +0.038 & 1 \\
2216.73 & 34.23 & $-$0.068 & +0.040 & 1 \\
2222.75 & 40.25 & $-$0.066 & +0.040 & 1 \\
2225.72 & 43.22 & $-$0.063 & +0.038 & 1 \\

2228.73 & 46.23 & $-$0.060 & +0.034 & 1 \\
2231.71 & 49.21 & $-$0.058 & +0.030 & 1 \\
2234.73 & 52.23 & $-$0.054 & +0.027 & 1 \\
2237.73 & 55.23 & $-$0.050 & +0.026 & 1 \\
2240.73 & 58.23 & $-$0.044 & +0.026 & 1 \\

2243.71 & 61.21 & $-$0.040 & +0.026 & 1 \\
2246.78 & 64.28 & $-$0.034 & +0.027 & 1 \\

\enddata
\tablenotetext{a} {The following corrections are to be $added$
to the data in Table 3 to correct them to the filter system
of Bessell (1990). Values in square brackets are less certain;
we have assumed that the YALO $B$-band filter corrections are 
represented by a flat function at the start. $t$ is the number of
days since the time of $B$-band maximum, JD 2,452,182.5.
$^b$YALO = 1; CTIO 0.9-m = 2, CTIO 1.5-m = 3.}
\end{deluxetable}

%

\begin{deluxetable}{cccccl}
\rotate
\tablewidth{0pc}
\tablecaption{Near-Infrared Photometry of SN 2001el$^a$}
\tablehead{   \colhead{JD$-$2,450,000} & \colhead{UT Date$^b$} &
\colhead {$J$} & \colhead{$H$} &
\colhead{$K$} & \colhead{Observer+Telescope} }
\startdata

2171.84 &  Sep19.34 & 13.607 (0.021) & 13.661 (0.036) & 13.543 (0.044) &  D. Gonzalez + YALO \\                         
2174.80 &  Sep22.30 & 13.129 (0.020) & 13.161 (0.037) & 13.021 (0.045) &  Espinoza + YALO \\            
2177.82 &  Sep25.33 & 12.878 (0.022) & 13.085 (0.037) & 12.831 (0.046) &  Espinoza + YALO \\                         
2183.83 &  Oct01.33 & 12.986 (0.020) & 13.193 (0.036) & 12.919 (0.043) &  D. Gonzalez + YALO \\                        
2186.79 &  Oct04.29 & 13.177 (0.020) & 13.253 (0.036) & 13.061 (0.043) &  Espinoza + YALO \\                         
2189.82 &  Oct07.33 & 13.538 (0.020) & 13.239 (0.036) & 13.116 (0.043) &  Espinoza + YALO \\                         
2192.82 &  Oct10.32 & 14.139 (0.022) & 13.276 (0.042) & 13.142 (0.051) &  D. Gonzalez + YALO \\                         
2195.74 &  Oct13.24 & 14.265 (0.032) & 13.245 (0.051) & 13.085 (0.063) &  D. Gonzalez/Pizarro + YALO \\          
2198.79 &  Oct16.29 & 14.290 (0.068) & 13.140 (0.099) & 13.036 (0.103) &  D. Gonzalez + YALO \\                        
2201.81 &  Oct19.31 & 14.380 (0.020) & 13.045 (0.036) & 12.987 (0.043) &  Pizarro/Espinoza + YALO \\          
2202.75 &   Oct20.25 &   14.191  (0.023) &    12.978  (0.040) &   13.035  (0.036) & S. Gonzalez/Krisciunas + LCO 1-m \\
2203.74 &   Oct21.24 &   14.167  (0.022) &    12.982  (0.019) &   12.957  (0.039) & S. Gonzalez + LCO 1-m \\
2204.70 &   Oct22.20 &   14.174  (0.025) &    12.966  (0.026) &   12.988  (0.036) & S. Gonzalez + LCO 1-m \\
2204.77 &  Oct22.27 & 14.263 (0.043) & 13.084 (0.071) & 12.873 (0.086) &  Espinoza + YALO \\                         
2207.84 &  Oct25.34 & 14.217 (0.021) & 13.061 (0.038) & 12.941 (0.045) &  Pizarro + YALO \\                        
2210.81 &  Oct28.31 & 14.199 (0.025) & 13.108 (0.042) & 13.045 (0.053) &  Pizarro + YALO \\                        
2213.67 &  Oct31.17 & 13.997 (0.023) & 13.132 (0.041) & 13.082 (0.049) &  Espinoza + YALO \\                         
2216.72 &  Nov03.22 & 14.112 (0.021) & 13.259 (0.039) & 13.317 (0.043) &  Espinoza + YALO \\                         
2222.74 &  Nov09.24 & 14.774 (0.025) & 13.769 (0.043) & 13.867 (0.058) &  D. Gonzalez + YALO \\                         
2225.71 &  Nov12.21 & 14.917 (0.028) & 13.812 (0.048) & 13.893 (0.058) &  D. Gonzalez + YALO \\                        
2228.72 &  Nov15.22 & 15.204 (0.026) & 13.991 (0.044) & 14.112 (0.053) &  Espinoza + YALO \\                         
2231.70 &  Nov18.20 & 15.354 (0.027) & 14.090 (0.045) & 14.305 (0.056) &  Espinoza + YALO \\                         
2234.72 &  Nov21.22 & 15.544 (0.026) & 14.213 (0.044) & 14.404 (0.059) &  D. Gonzalez + YALO \\                        
2237.73 &  Nov24.23 & 15.799 (0.033) & 14.349 (0.053) & 14.537 (0.065) &  D. Gonzalez + YALO \\                        
2238.79 &   Nov25.29 &   15.713  (0.020) &    14.355  (0.054) &   14.711  (0.038) & S. Gonzalez + LCO 1-m \\
2240.70 &   Nov27.20 &   15.879  (0.020) &    14.514  (0.039) &   14.690  (0.053) & S. Gonzalez + LCO 1-m \\
2240.72 &  Nov27.22 & 15.974 (0.034) & 14.488 (0.055) & 14.711 (0.077) &  D. Gonzalez + YALO \\                        
2241.69 &   Nov28.19 &   15.964  (0.020) &    14.492  (0.016) &   14.798  (0.048) & S. Gonzalez + LCO 1-m \\
2242.69 &   Nov29.19 &   16.030  (0.025) &    14.562  (0.022) &   14.751  (0.059) & S. Gonzalez + LCO 1-m \\
2243.68 &   Nov30.18 &   16.069  (0.024) &    14.548  (0.080) &                   & S. Gonzalez + LCO 1-m \\
2243.74 &  Nov30.24 & 16.238 (0.033) & 14.576 (0.055) & 14.780 (0.075) &  Espinoza + YALO \\                         
2244.71 &  Dec01.21 &   16.089  (0.046) &                    &   14.833  (0.044) & S. Gonzalez + LCO 1-m \\
2246.77 &  Dec03.27 & 16.420 (0.043) & 14.824 (0.067) & 14.938 (0.081) &  Espinoza + YALO \\              
\enddata
\tablenotetext{a} {The data are based on differential photometry with respect
to star 6 given in Table 2. The YALO data are presented {\em without} the corrections 
given in Table 6.
$^b$Year is 2001.}
\end{deluxetable}

\begin{deluxetable}{ccccc}
\tablewidth{0pc}
\tablecaption{Corrections to YALO Infrared Photometry$^a$}
\tablehead{   \colhead{JD$-$2,450,000} & \colhead{$t$ (d)} &
\colhead{$\Delta J$} &
\colhead{$\Delta H$} & \colhead{$\Delta K$} }
\startdata
2171.84 & $-$10.66 & 0.030 & 0.011 & $-$0.011 \\
2174.80 & $-$7.70  & 0.042 & 0.004 &  0.000 \\
2177.82 & $-$4.68  & 0.051 & $-$0.003 & 0.010 \\
2183.83 & 1.33     & 0.071 & $-$0.016 & 0.032 \\
2186.79 & 4.29     & 0.041 & $-$0.038 & 0.056 \\

2189.82 & 7.32  & 0.014 & $-$0.042 &  0.054 \\
2192.82 & 10.32 & $-$0.013 & $-$0.044 & 0.044 \\
2195.74 & 13.24 & $-$0.037 & $-$0.045 & 0.033 \\
2198.79 & 16.29 & $-$0.062 & $-$0.049 & 0.021 \\
2201.81 & 19.31 & $-$0.080 & $-$0.059 & 0.010 \\

2204.77 & 22.27 & $-$0.102 & $-$0.070 & $-$0.001 \\
2207.84 & 25.34 & $-$0.070 & $-$0.071 & $-$0.012 \\
2210.81 & 28.31 & $-$0.051 & $-$0.062 & $-$0.016 \\
2213.67 & 31.17 & $-$0.082 & $-$0.022 & $-$0.009 \\
2216.72 & 34.22 & $-$0.097 & $-$0.005 & $-$0.007 \\

2222.74 & 40.24 & $-$0.125 & 0.030 &  0.004 \\
2225.71 & 43.21 & $-$0.131 & 0.033 &  0.000 \\
2228.72 & 46.22 & $-$0.131 & 0.033 &  0.000 \\
2231.70 & 49.20 & $-$0.131 & 0.033 &  0.000 \\
2234.72 & 52.22 & $-$0.131 & 0.033 &  0.000 \\

2237.73 & 55.23 & $-$0.131 & 0.033 &  0.000 \\
2240.72 & 58.22 & $-$0.131 & 0.033 &  0.000 \\
2243.74 & 61.24 & $-$0.131 & 0.033 &  0.000 \\
2246.77 & 64.27 & $-$0.131 & 0.033 &  0.000 \\

\enddata
\tablenotetext{a} {The following corrections are to be $added$
to the YALO data in Table 5 to correct them to the photometric system
of Persson et al. (1998). $t$ is the number of days since
the time of $B$-band maximum, JD 2,452,182.5.} 
\end{deluxetable}

\begin{deluxetable}{cccc}
\tablewidth{0pc}
\tablecaption{Maximum Magnitudes of SN 2001el}
\tablehead{   \colhead{Filter} &
\colhead{JD$-$2,450,000} & \colhead{m$_{max}$ } }
\startdata
$U$   &  2181.4 (1.0) & 12.67 (0.04) \\
$B$   &  2182.5 (0.5) & 12.81 (0.02) \\
$V$   &  2184.5 (0.5) & 12.73 (0.02) \\
$R$   &  2185.3 (0.5) & 12.62 (0.04) \\
$I$   &  2179.1 (1.0) & 12.82 (0.04) \\
$J$   &  2178.4 (1.0) & 12.90 (0.04) \\
$H^a$   &  2177.8 (1.0) & 13.08 (0.04) \\
$K$   &  2178.8 (1.0) & 12.83 (0.04) \\
\enddata
\tablenotetext{a} {The second maximum 
($H$ = 12.97 at JD 2,452,204.7) is actually 
brighter than the first $H$-band maximum.}
\end{deluxetable}

\begin{deluxetable}{cccc}
\tablewidth{0pc}
\tablecaption{Color Excesses and Implied Extinction$^a$}
\tablehead{   \colhead{Color Index} &
\colhead{Color Excess} & \colhead{A$_V$ (host galaxy)} &
\colhead{A$_V$ (total)} }
\startdata
($B-V$)$_{max}$ & 0.185 (0.104) & 0.53 (0.30) & 0.57 (0.30) \\
($B-V$)$_{tail}$ & 0.253 (0.063) & 0.73 (0.19) & 0.77 (0.19) \\
($V-I$)$_{max}$  & 0.133 (0.087) & 0.31 (0.20) & 0.35 (0.20) \\
($B-V$)$_{avg}$ & 0.206 (0.046) & 0.593 (0.136) & 0.636 (0.136) \\
$V-J$ & 0.350 (0.077) & 0.445 (0.114) & 0.488 (0.114) \\
$V-H$ & 0.450 (0.071) & 0.502 (0.089) & 0.545 (0.089) \\
$V-K$ & 0.530 (0.054) & 0.554 (0.063) & 0.597 (0.063) \\

%
%

\enddata
\tablenotetext{a} {All values are measured in magnitudes.
Values in parentheses are 1-$\sigma$ uncertainties.
We adopt R$_V \equiv $ A$_V$ / E($B-V$) = 2.88 $\pm$ 0.15  (Wang et al.
2002) for the host galaxy. We also adopt E($B-V$) = 0.8 E($V-I$).}
\end{deluxetable}

\begin{deluxetable}{lcccc}
\tablewidth{0pc}
\tablecaption{Wavelength Shifts to Instrumental Bandpasses$^a$}
\tablehead{   \colhead{Telescope} &
\colhead{$B$} & \colhead{$V$} & \colhead {$R$} &
\colhead{$I$} }
\startdata
YALO       & 25 red  & 100 blue & 30 red  & 100 blue \\
CTIO 0.9-m & 25 red  & 0        & 50 blue & 50 red \\
CTIO 1.5-m & 25 blue & 0        & 0       & 0  \\
\enddata
\tablenotetext{a} {All values are measured in \AA.}
\end{deluxetable}

\clearpage

\begin{figure}
\figurenum{1}
\epsscale{0.5}
\plotone{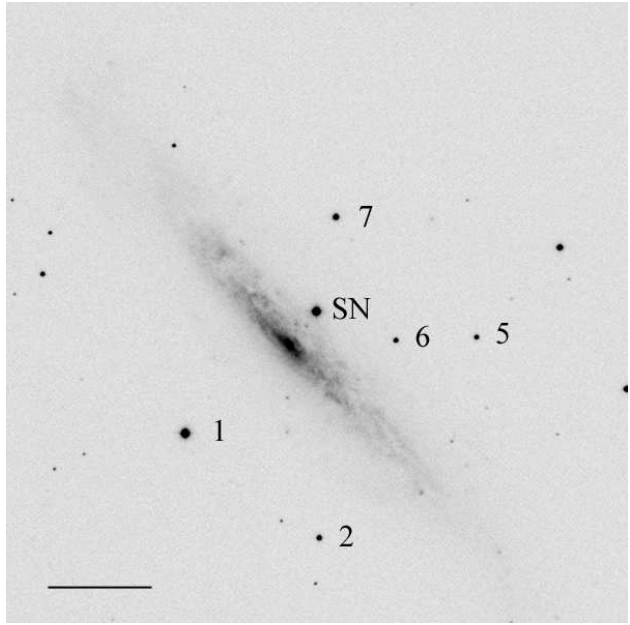}
\caption{A $V$-band image of SN 2001el in NGC 1448 obtained with the CTIO 0.9-m
telescope on 26.4 Sept 2001. The exposure time was 25 seconds. The
local photometric standards are shown. The bar corresponds to 1   
arcminute. North is up and East is to the left. The supernova is  
marked as ``SN''.  
}
\end{figure}

\clearpage

\begin{figure}
\figurenum{2} 
\epsscale{0.8}
\plotone{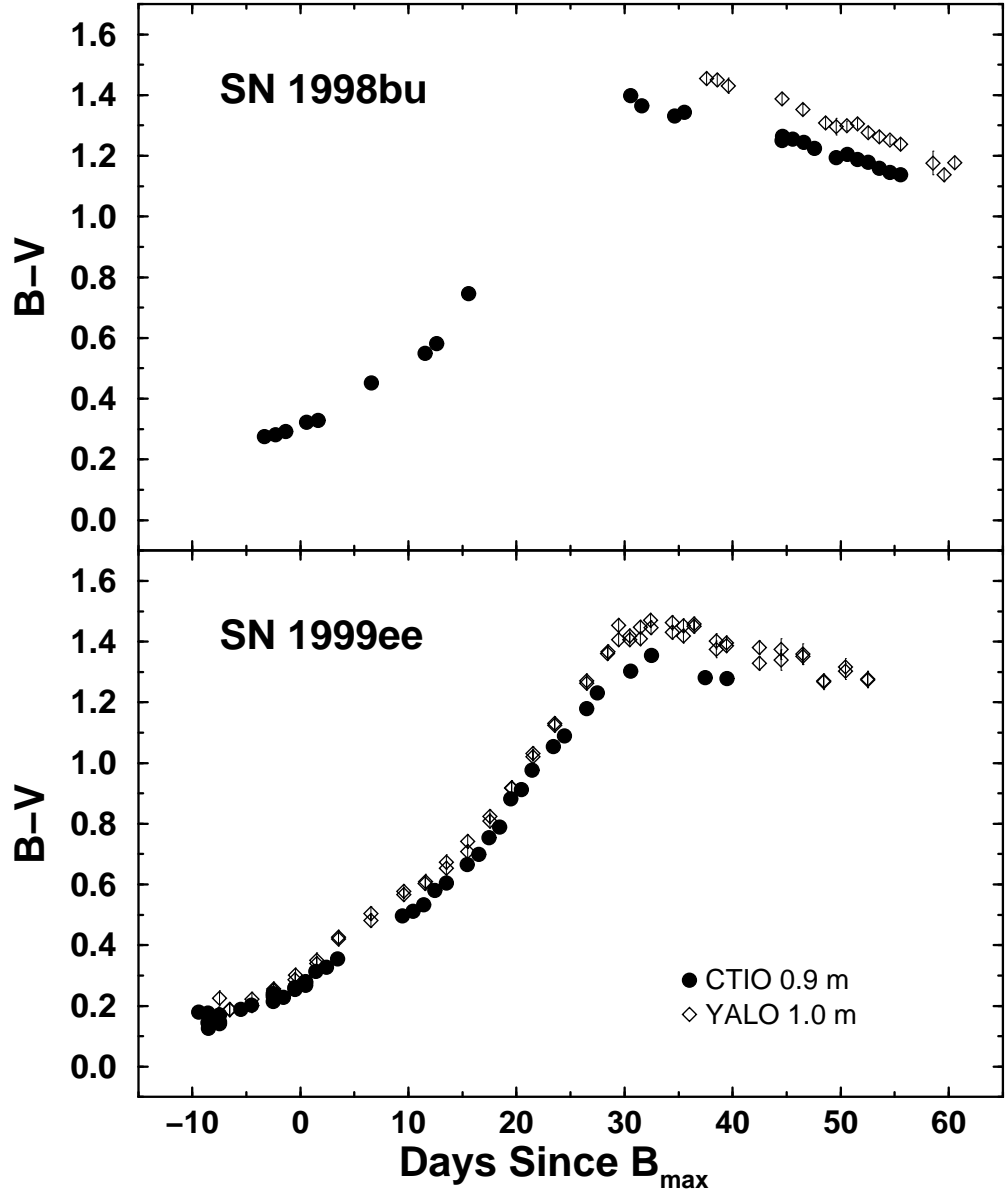}
\caption{$B-V$ colors of SNe 1998bu and 1999ee, as observed with
the CTIO 0.9-m and YALO telescopes.  Due to differences in the
filter transmission curves, late-time $B-V$ colors derived from YALO
data are roughly 0.12 mag redder.}
\end{figure}

\begin{figure}
\figurenum{3} 
\epsscale{0.8}
\plotone{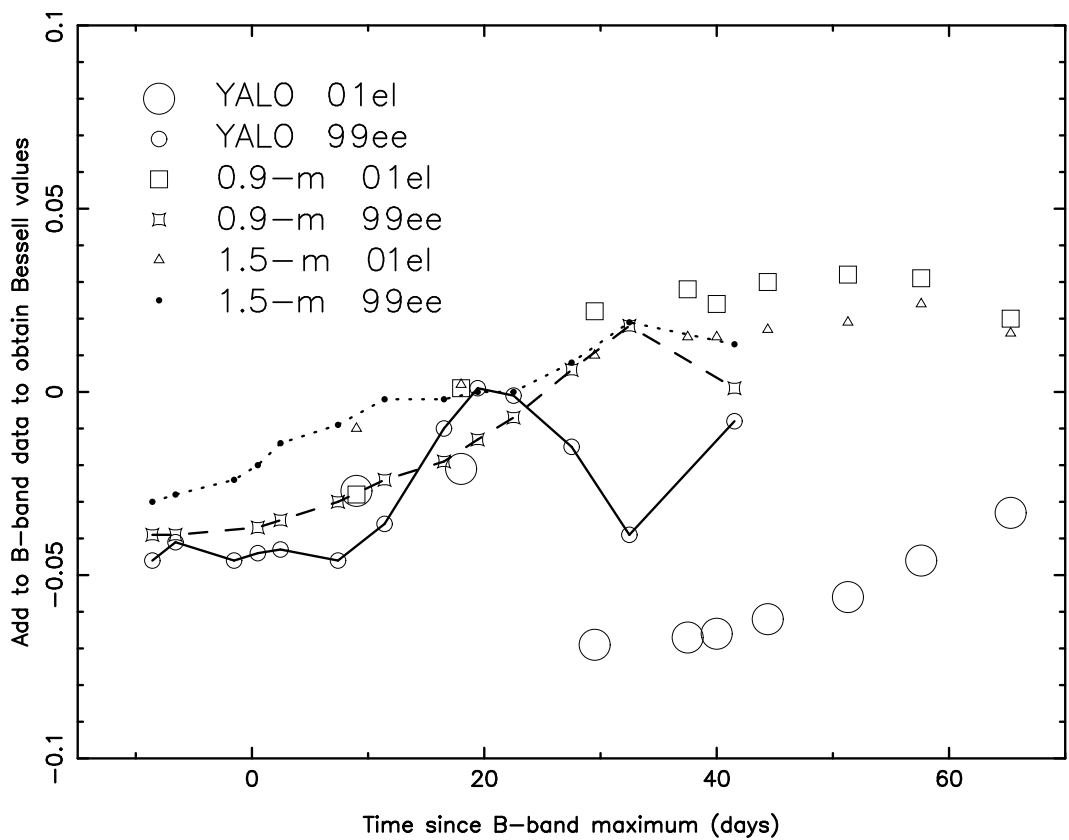}
\vspace {5 mm}
\caption{Based on synthetic photometry using Bessell (1990), YALO,
CTIO 0.9-m, and CTIO 1.5-m filter functions, we show $B$-band corrections to
YALO and CTIO data to place photometry of SNe 1999ee and
2001el on the Bessell system.  The time of $B$-band maximum is
JD 2,451,469.1 for SN 1999ee (Stritzinger et al. 2002) and 2,452,182.5
for SN 2001el.  We used spectra of SN 1999ee given
by Hamuy et al. (2002), and spectra of SN 2001el by
Wang et al. (2002) and Nugent (2002, private communication). 
The filter corrections for SN 1999ee are connected by lines, telescope by
telescope, to guide the eye.}
\end{figure}

\begin{figure}
\figurenum{4} 
\epsscale{0.8}
\plotone{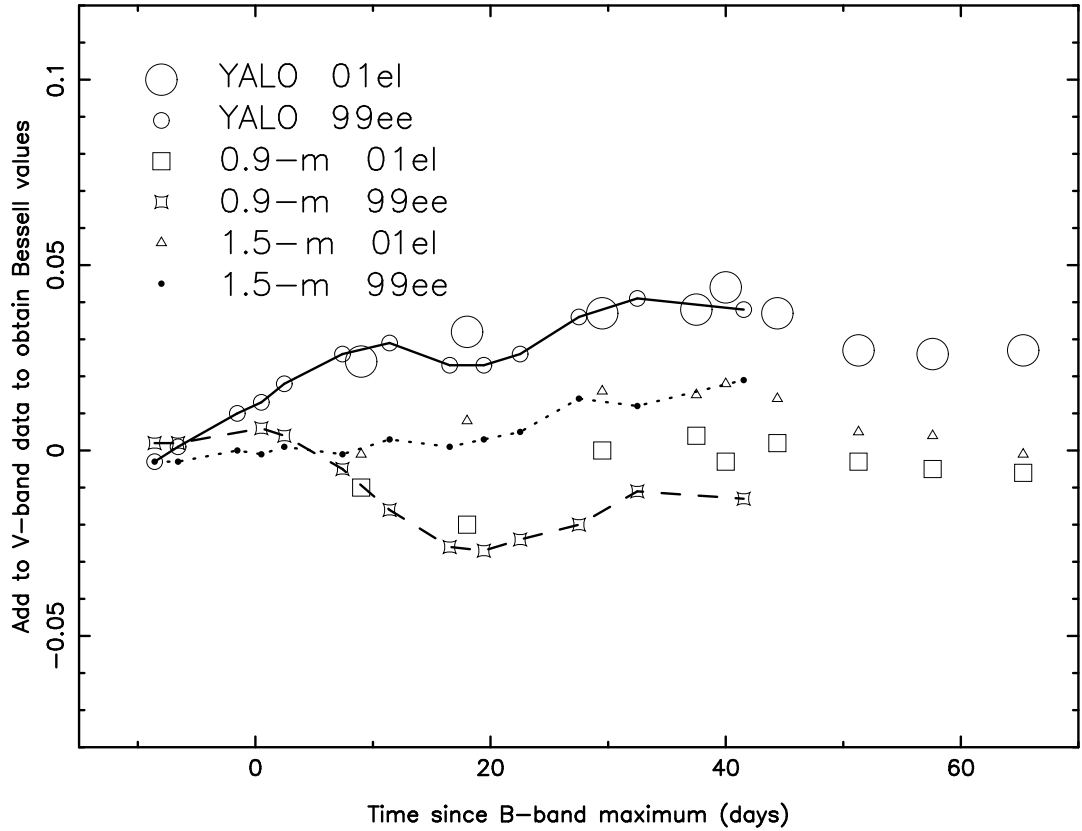}
\vspace {5 mm}
\caption{Same as Fig. 3, but for the $V$-band.  Note that, unlike the $B$-band,
these corrections are mostly telescope-dependent, not telescope- and
object-dependent after $t$ = 20 d.}
\end{figure}

\begin{figure}
\figurenum{5}
\epsscale{0.8}
\plotone{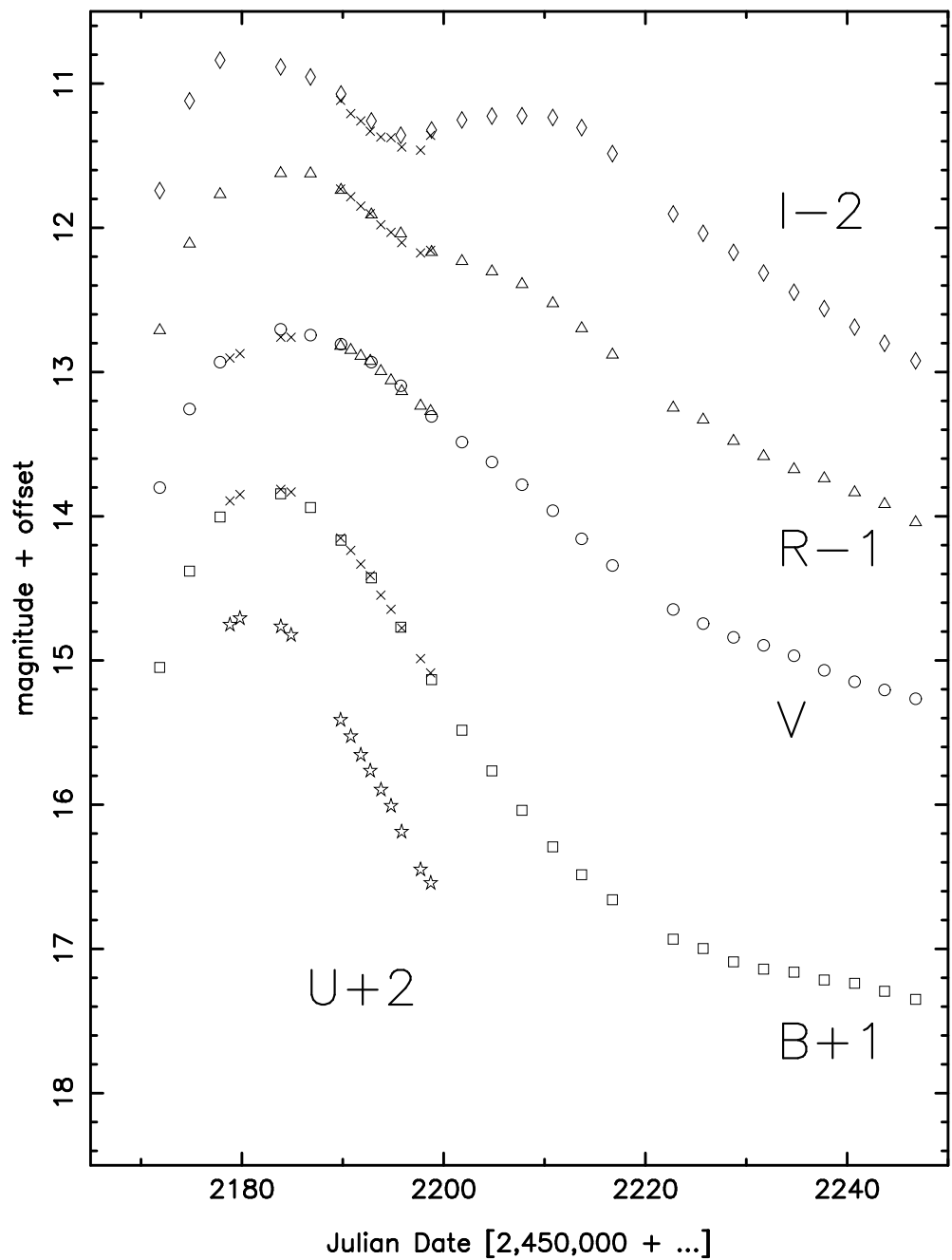}
\caption{Optical photometry of SN 2001el.  For $B$, $V$, $R$, and $I$ the
YALO data are plotted as smaller open symbols, while data from the CTIO
0.9-m and 1.5-m telescopes are plotted as $\times$'s.
We have corrected  the $B$-band and $V$-band data to the filter system of
Bessell (1990) using the values in Table 4.
On the whole the internal errors are smaller than
the size of the points.  The $U$, $B$, $R$, and $I$
data have been offset vertically by +2, +1, $-$1, and $-$2,
magnitudes, respectively. The Landolt data are plotted as nightly means.
Data from two nights are off the right edge of the plot.}
\end{figure}

\clearpage

\clearpage 

\begin{figure}
\figurenum{6}
\epsscale{0.8}
\plotone{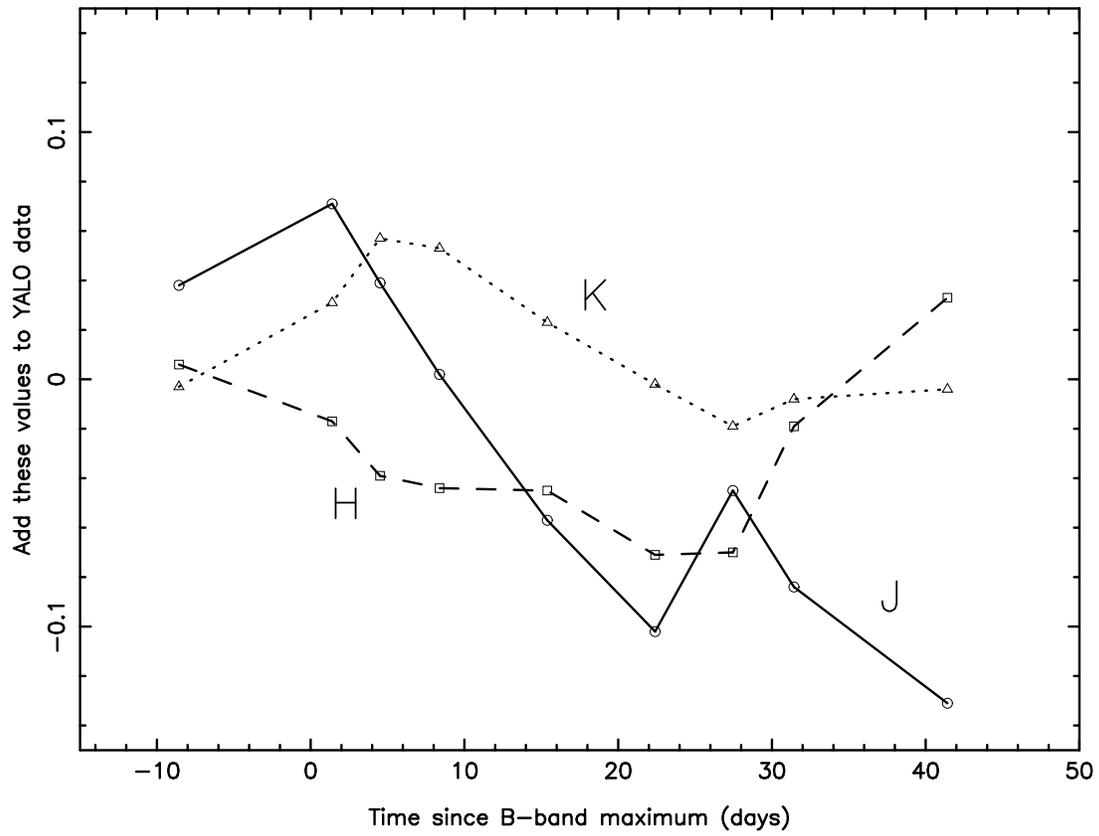}
\vspace {5 mm}
\caption{Corrections to YALO infrared photometry, based on spectra
of SN 1999ee (Hamuy et al. 2002), to place the YALO data on the
system of Persson et al. (1998).  Solid line = $J$; dashed line = $H$;
dotted line = $K$.}
\end{figure}

\clearpage 

\begin{figure}
\figurenum{7}
\epsscale{0.8}
\plotone{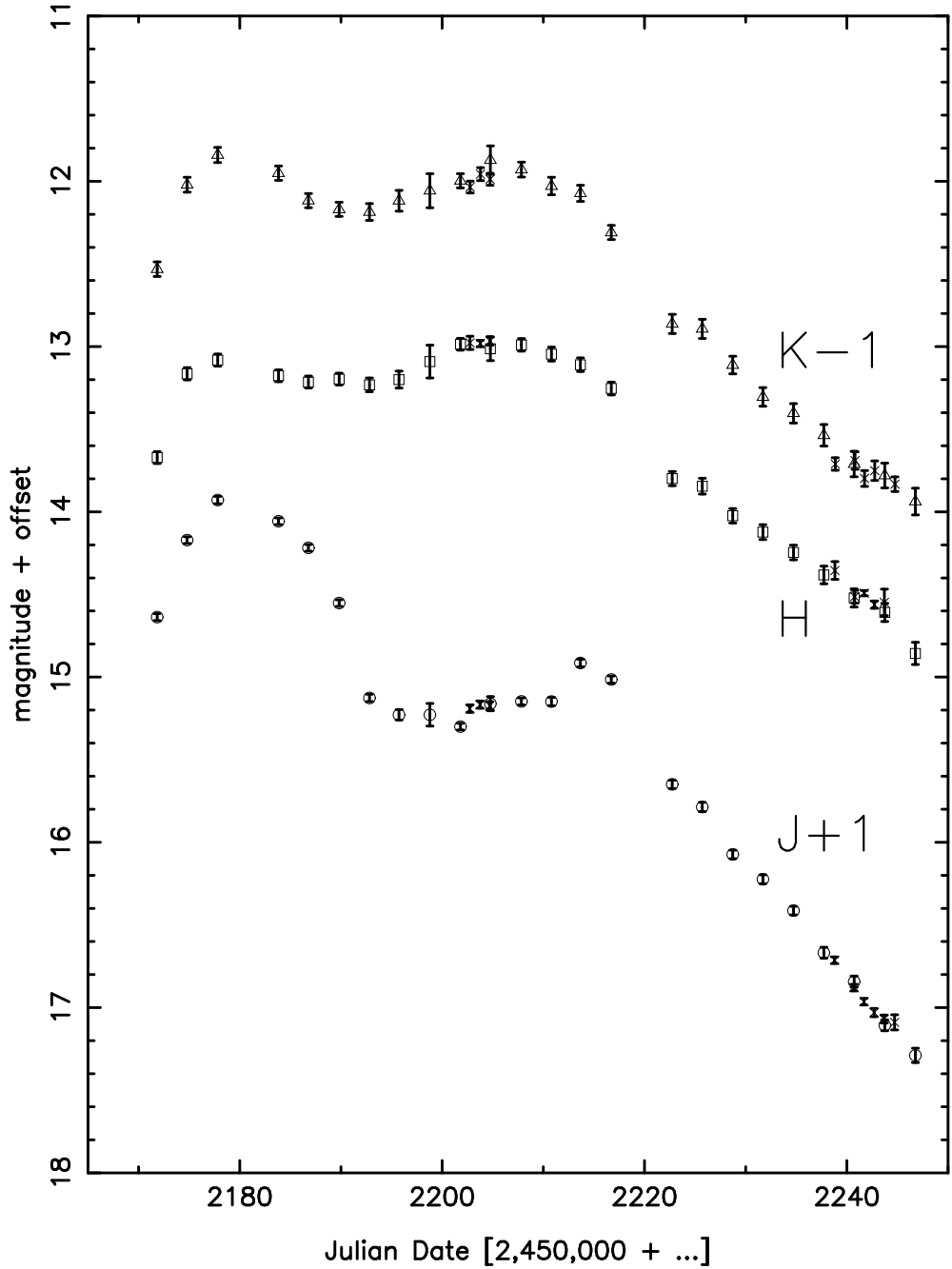}
\caption{$JHK$ photometry of SN 2001el. The $J$ and $K$ data have
been offset vertically by +1 and $-$1 magnitudes, respectively. 
Open symbols = YALO data. $\times$'s = LCO 1-m data.  The
YALO data include the corrections given in Table 6.}
\end{figure}

\clearpage 

\begin{figure}
\figurenum{8}
\epsscale{0.8}
\plotone{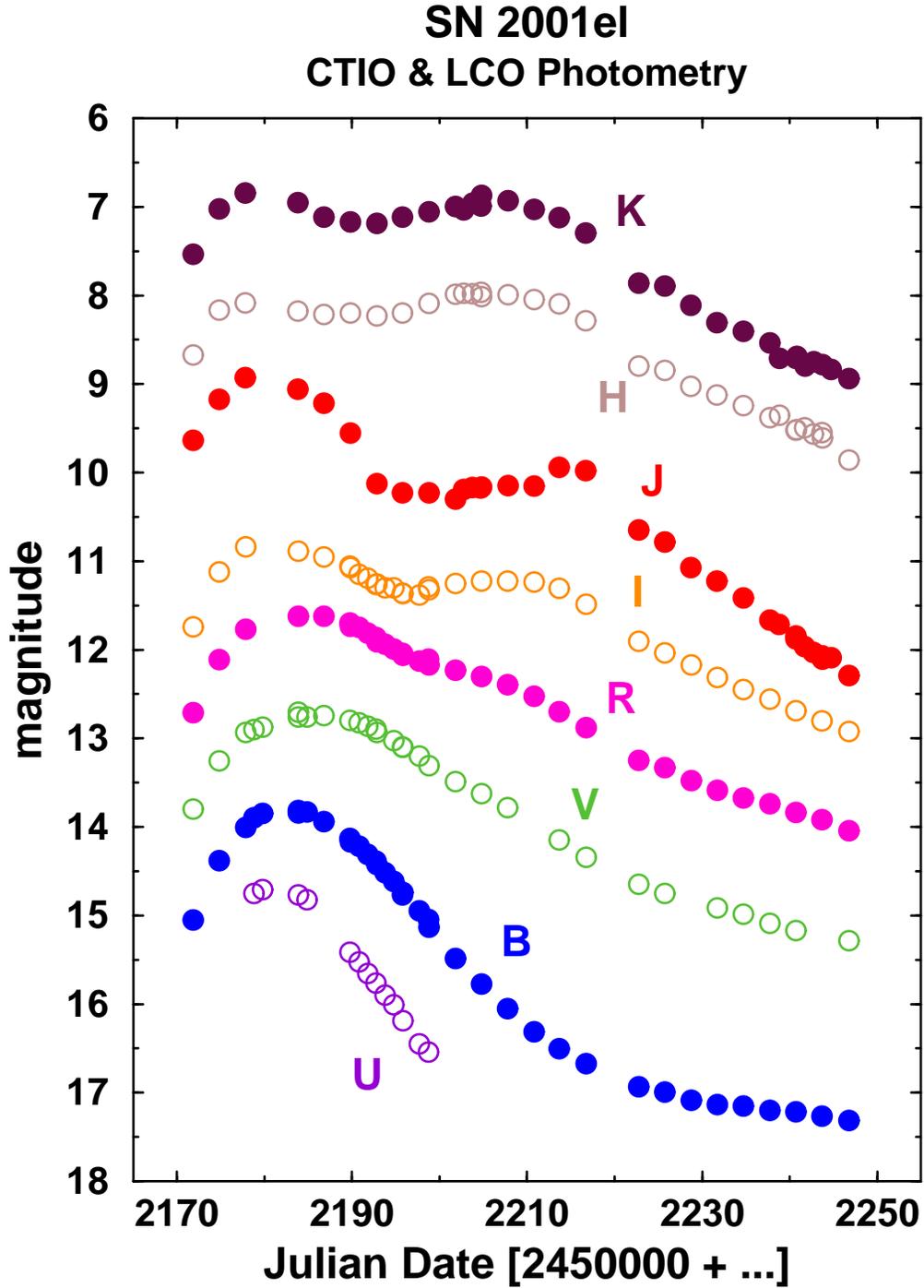}
\caption{$UBVRIJHK$ light curve of SN 2001el.  The different
bands are offset for plotting purposes but show that the
red and infrared maxima occurred earlier than for the $UBV$ bands.}
\end{figure}

\begin{figure}
\figurenum{9}
\epsscale{0.8}
\plotone{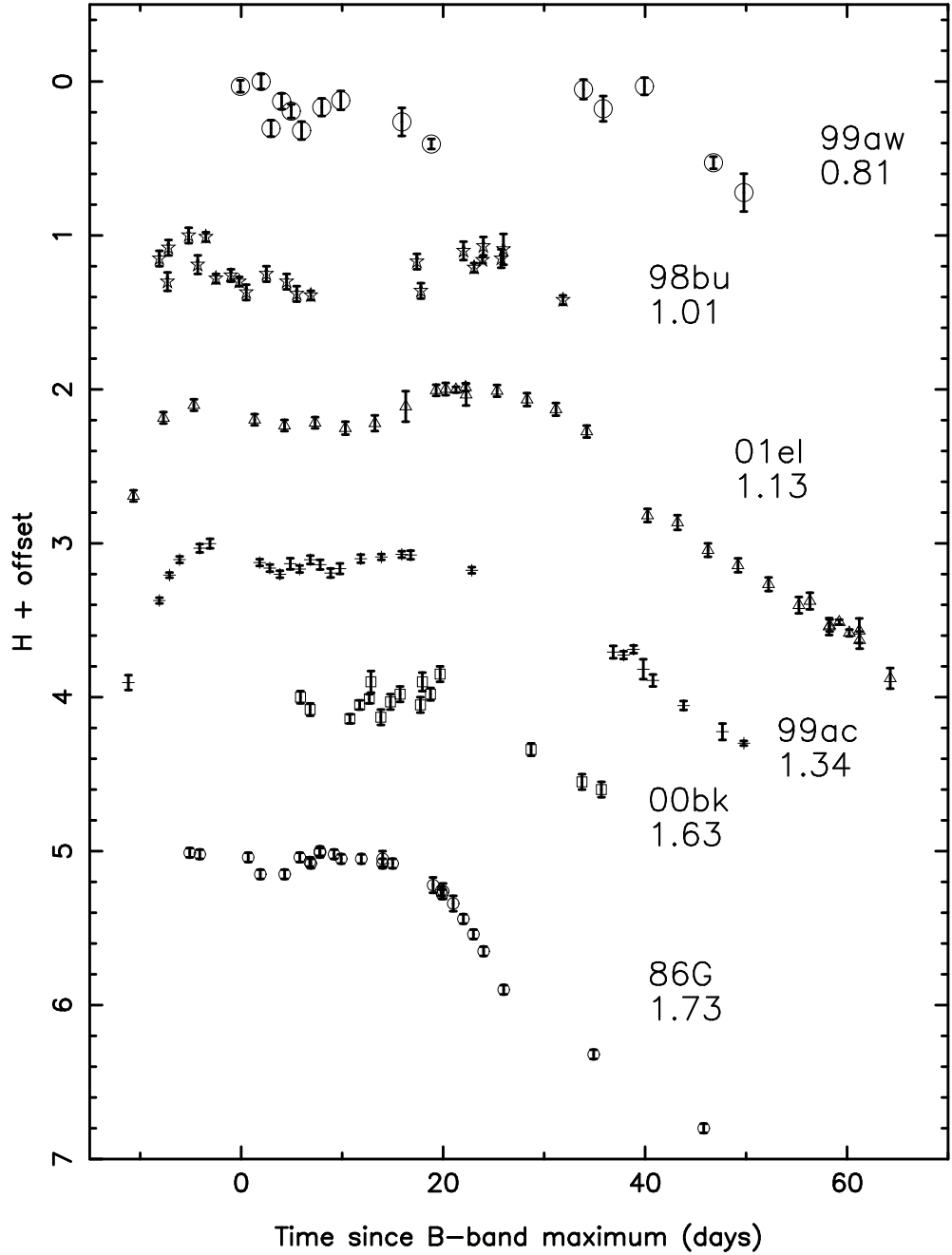}
\caption{$H$-band light curves of a number of Type Ia supernovae.
The objects are ordered from top to bottom by the decline rate parameter
$\Delta$m$_{15}$(B).
Sources: SN 1999aw, Strolger et al. (2002); SN 1998bu, see Meikle
(2000); SN 2001el (this paper); SN 1999ac, Phillips et al. (2002);
SN 2000bk (Krisciunas et al. 2001); SN 1986G (Frogel et al. 1987).}
\end{figure}

\begin{figure}
\figurenum{10}
\epsscale{0.8}
\plotone{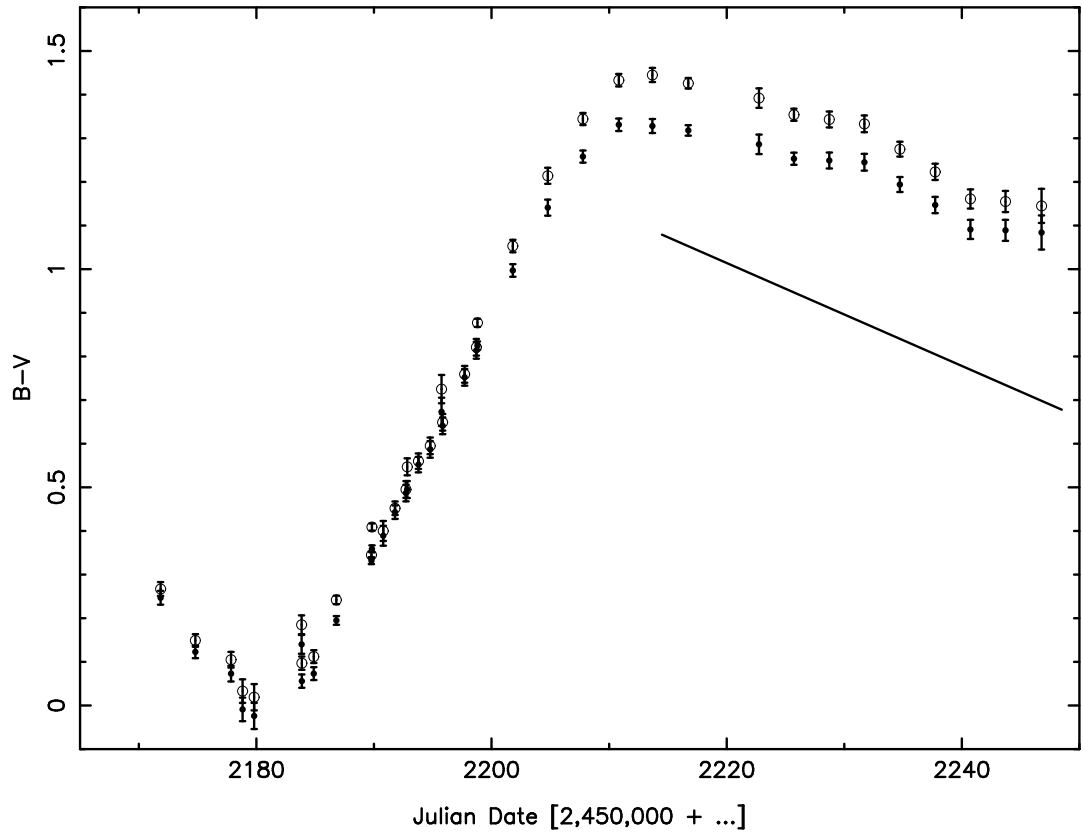}
\vspace {5 mm}
\caption{$B-V$ colors of SN 2001el versus Julian Date.  
The smaller open circles represent the uncorrected data, while the
solid dots include the corrections of Table 4.  
The solid line is the zero reddening line of
Lira (1995) adjusted to the time of $V$-band maximum given in Table 7.}
\end{figure}

\begin{figure}
\figurenum{11}
\epsscale{0.8}
\plotone{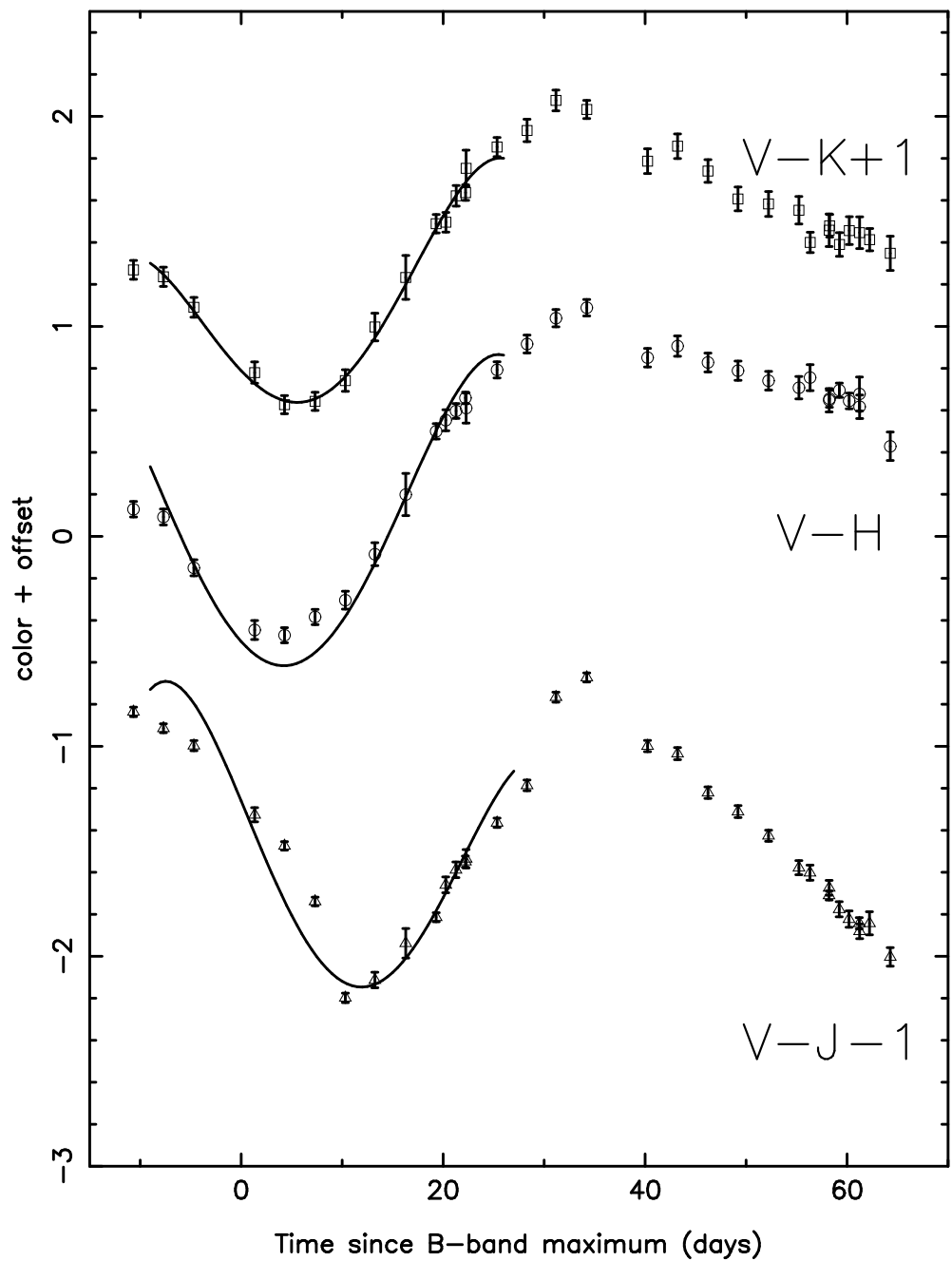}
\caption{V {\em minus} near-IR color curves for SN 2001el, using the
$V$-band data of Table 3, corrected by the values in Table 4,
and the $JHK$ data in Table 5, with the YALO
data corrected according the values in Table 6. The V$-$K
data have been offset by +1 mag, while the $V-J$ data have
been offset by $-$1 mag.  The solid lines are based on the
fourth order polynomial fits to data of eight Type Ia SNe studied by
Krisciunas et al. (2000) which have mid-range B-band decline rates. The
loci are adjusted in the ordinate direction to minimize the reduced
$\chi^2$ of the fits.}
\end{figure}

\begin{figure}
\figurenum{12}
\epsscale{0.8}
\plotone{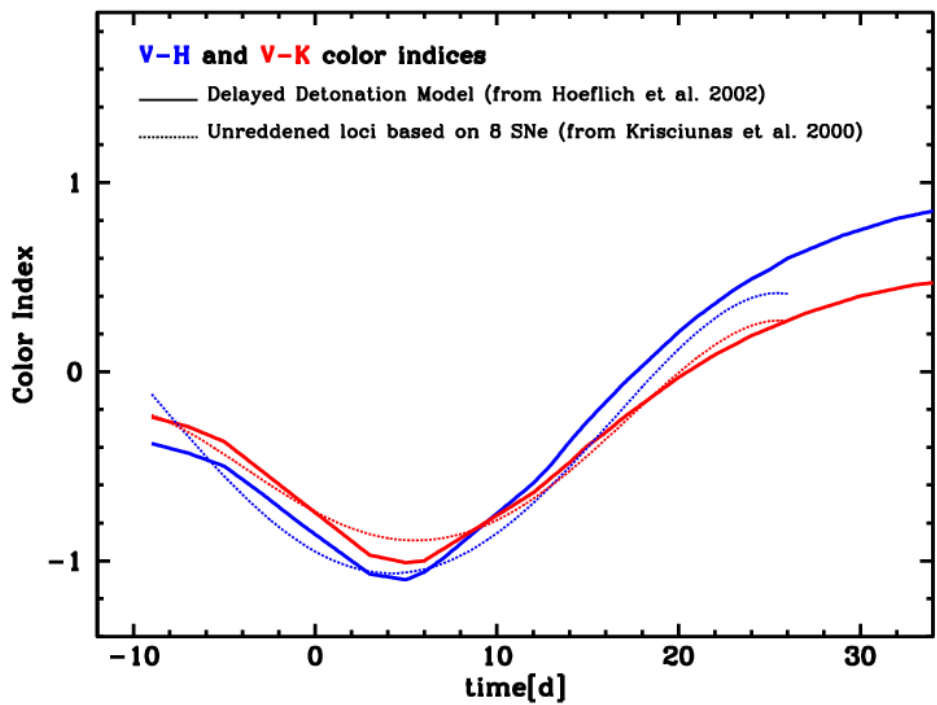}
\caption{Color curves based on a 
delayed detonation model of a typical Type Ia SN
(H\"{o}flich et al. 2002). The abscissa is the number of
days since $B$-band maximum.  Model uncertainties allow the
theoretical curves to be shifted $\pm$1 d in the $x$-direction.
$V-H$ loci are blue lines
while $V-K$ loci are red. Also shown (as dotted lines) are the 
unreddened loci based on 8 mid-range decliners studied by 
Krisciunas et al. (2000). Given the 0\fm2 to 0\fm3 accuracy of 
the infrared models, the agreement between theory and observation
is better than expected. }
\end{figure}

\begin{figure}
\figurenum{13}
\epsscale{0.8}
\plotone{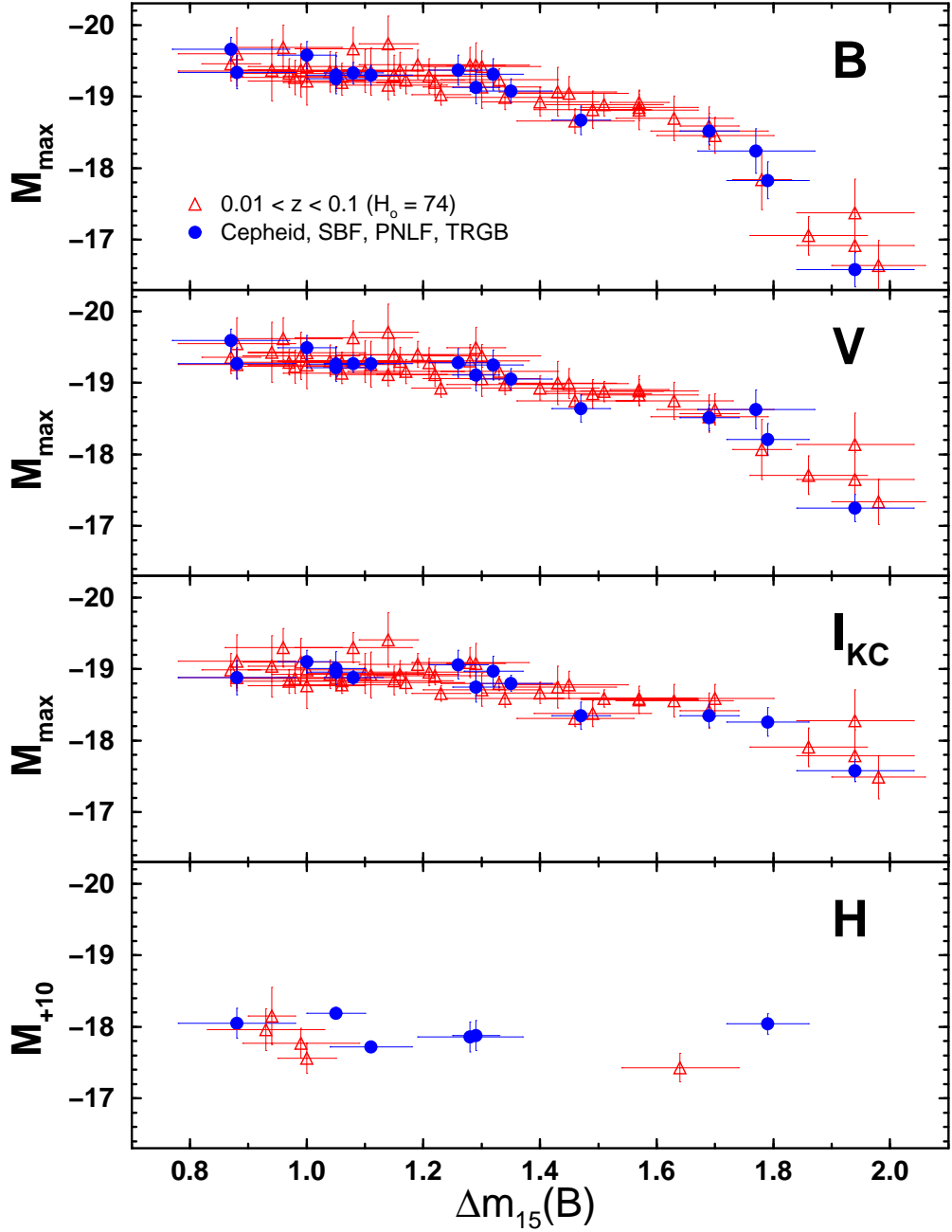}
\caption{Absolute magnitudes of Type Ia SNe versus decline rate
parameter $\Delta$m$_{15}$(B).  For the $H$-band the absolute
magnitudes correspond to 10 days after the time of $B_{max}$.
The solid points were derived using
distances from Cepheids or other direct measures of the
distance to the host galaxy such as surface brightness fluctuations,
the planetary nebula luminosity function, or using
the ``tip of the red giant branch method''.  Open triangles correspond
to points derived using a Hubble constant of 74 km s$^{-1}$ Mpc$^{-1}$;
these objects are sufficiently distant to be in the Hubble flow.
The right-most point in the bottom panel corresponds to SN 1986G,
which was highly reddened and whose distance is subject to some
uncertainty.}
\end{figure}

\end{document}